\documentclass[%
 reprint,
 amsmath,amssymb,
 aps,
floatfix,
]{revtex4-2}

\usepackage{graphicx}
\usepackage{dcolumn}
\usepackage{bm}
\usepackage{hyperref}
\usepackage{braket}
\usepackage{dsfont}
\usepackage{color}
\usepackage{soul}

\newcommand{\trace}[0]{\operatorname{Tr}}
\newcommand{\tr}[0]{\operatorname{Tr}}

\newcommand{\be}{\begin{equation}}
\newcommand{\ee}{\end{equation}}
\newcommand{\ba}{\begin{array}}
\newcommand{\ea}{\end{array}}
\newcommand{\bqa}{\begin{eqnarray}}
\newcommand{\eqa}{\end{eqnarray}}

\begin{document}

\preprint{APS/123-QED}

\title{Efficient Assessment of Process Fidelity}

\author{Sean Greenaway}
\author{Fr{\'e}d{\'e}ric Sauvage}
\author{Kiran E. Khosla}
\author{Florian Mintert}
\affiliation{Physics Department, Blackett Laboratory, Imperial College London, Prince Consort Road, SW7 2BW, United Kingdom}

\date{\today}

\begin{abstract}

The accurate implementation of quantum gates is essential for the realisation of quantum algorithms and digital quantum simulations. This accuracy may be increased on noisy hardware through the variational optimisation of gates, however the experimental realisation of such a protocol is impeded by the large effort required to estimate the fidelity of an implemented gate.
With a hierarchy of approximations we find a faithful approximation to the quantum process fidelity that can be estimated experimentally with reduced effort.
Its practical use is demonstrated with the optimisation of a three-qubit quantum gate on a commercially available quantum processor.

\end{abstract}

\maketitle

\section{Introduction}

Experimental progress in developing quantum computers has led to the realisation of noisy intermediate scale quantum (NISQ) devices in a wide array of experimental platforms~\cite{friis2018observation,wang2018efficient,cervera2018exact,otterbach2017unsupervised}. Whilst the number of qubits in these devices is approaching that needed for quantum supremacy~\cite{arute2019quantum}, their noisiness remains a fundamental limiting factor in the development of useful applications~\cite{preskill2018quantum}. As such, a number of techniques have been developed for \textit{error mitigation} in quantum computations, wherein additional measurement data and classical post-processing are used in order to extract relatively noise-free results from the noisy devices~\cite{mcardle2019error,sagastizabal2019experimental,zhang2020error,vovrosh2021efficient}.

Many of these techniques have focused on obtaining accurate expectation values from noisy devices~\cite{temme2017error,kandala2019error,li2017efficient,endo2018practical,he2020resource,dumitrescu2018cloud} and are most commonly paired with variational quantum algorithms (VQAs)~\cite{mcclean2016theory,higgott2019variational,jones2019variational,khatri2019quantum,larose2019variational,cirstoiu2019variational}, hybrid quantum-classical algorithms in which a parameterised ansatz is updated using experimental measurements in order to optimise for some relevant observable. An additional, complementary error mitigation strategy which could be immensely powerful would be to variationally optimise quantum channels by adapting them from their textbook implementation such that the optimised channel more closely realises the desired operation~\cite{heya2018variational}.

For optimal implementations of such optimisations it is helpful to have the ability to efficiently assess the accuracy of a quantum channel and to distinguish imperfections in state preparation and in read-out from imperfections in the actual quantum dynamics. The latter aspect can be addressed in terms of randomised benchmarking~\cite{carignan2015characterizing,helsen2020general,harper2020efficient} that can be applied to standard protocols for channel evaluation~\cite{cai2019constructing,erhard2019characterizing}. The goal of this paper is to address the former aspect, {\it i.e.} the efficiency of fidelity assessment.

The process fidelity~\cite{gilchrist2005distance,nielsen2002simple} is very expensive to evaluate experimentally.
Instead of evaluating it directly, it is typically {\em estimated}.
Since this can be done in terms of a relatively small number of expectation values independent of the system size~\cite{flammia2011direct,da2011practical}, this far more efficient than an exact evaluation.
The fact that the estimation is unbiased with a relatively low variance is sufficient for channel optimisation: techniques from quantum optimal control which have been shown to be highly effective in similar settings~\cite{sauvage2020optimal,mukherjee2020bayesian,mukherjee2020preparation,kuros2020phase,goerz2014optimal} may be applied here.

Unfortunately, whilst estimates of the process fidelity are theoretically efficient, their implementation on real hardware necessitates changing the experimental setup at every shot of the experiment. With the limited speed at which NISQ hardware may be controlled and accessed, this translates to a substantial decrease in efficiency, preventing the estimation protocol from being usable in optimisation.

An alternative figure of merit which may be implemented using only a small number of unique experimental settings is thus desirable. In this work we introduce a hierarchy of approximations to the process fidelity which we refer to as {\em $k-$fidelities}. These are given in terms of a physically implementable set of expectation values which, together with the fact they are approximately monotonic functions of the process fidelity, means they have the potential to provide alternative figures of merit by which the quality of quantum channels may be assessed.

In particular, the leading order term, the $0-$fidelity, is especially useful since it satisfies the requirement of being efficiently estimable with few unique experimental settings. In order to keep the optimisation target the same, it is crucial that an approximation to a target function is maximised if and only if the target function is also maximised. Such approximations are known as \textit{faithful}~\cite{cerezo2021cost}. The $0-$fidelity is a faithful approximation to the process fidelity, making it a suitable figure of merit for the optimisation of quantum channels. We find that the $0-$fidelity not only approximates the process fidelity well, particularly at high fidelities, but we also find that the approximation becomes better as the system size increases, which we demonstrate numerically.

The key advantage of the $0-$fidelity over the process fidelity is that it can be efficiently estimated even under the constraints imposed by NISQ platforms, allowing the estimation protocol to be repeated multiple times as necessitated by an optimisation routine. We demonstrate the superior performance of the $0-$fidelity estimations under these conditions both numerically and through experiments performed on an IBM quantum device.

\section{The $\mathbf{0-}$Fidelity}

\subsection{Evaluating the Quality of Quantum Channels} \label{sec:quality_channels}

Any attempt to realise a desired channel $\Lambda$ that maps input states $\rho_I$ to their designated output states $\rho_f=\Lambda(\rho_I)$ experimentally will inevitably result in the realisation of a channel $\Gamma$ that does not perfectly coincide with $\Lambda$.
The similarity between $\Gamma$ and $\Lambda$ is typically quantified by the \textit{process fidelity},
\be
  F(\Lambda, \Gamma) = \frac{1}{d^2} \sum_{i=1}^{d^2} \tr[\Lambda(\sigma_i^\dagger)\Gamma(\sigma_i)]\ ,
  \label{eq:process_fidelity}
\ee
expressed in terms of a complete set of mutually orthonormal operators $\sigma_i$ on a $d$-dimensional Hilbert space.

In order to assess the process fidelity experimentally, it is essential that these inputs be quantum states, \textit{i.e.} Hermitian, positive semi-definite operators. However, while a complete set of operators contains $d^2$ elements, there are only $d$ mutually orthogonal quantum states and thus Eq.~\eqref{eq:process_fidelity} cannot be used directly and must be adapted such that the inputs are quantum states.

A formulation of the process fidelity that is consistent with the requirement that the channel inputs be quantum states is given by
\be
  F(\Lambda, \Gamma) = \frac{1}{d^2} \sum_{ij=1}^{d^2} [B^{-1}]_{ij} \tr[\Lambda(\rho_i)\Gamma(\rho_j)]\ ,
 \label{eq:general_process_fidelity}
\ee
with the matrix $B_{ij} = \tr[\rho^\dagger_i \rho_j]$ comprised of the mutual overlaps of the states $\rho_i$. The change in notation from $\sigma_i$ to $\rho_i$ emphasises the experimentally motivated restriction to quantum states.

The term $\tr[\Lambda(\rho_i)\Gamma(\rho_j)]$ in Eq.~\eqref{eq:general_process_fidelity} can be understood as the expectation value of the observable $\Lambda(\rho_i)$ with respect to the state $\Gamma(\rho_j)$, {\it i.e.} the state obtained with the evolution described by the channel $\Gamma$ after initialization in the state $\rho_j$. For most channels $\Lambda$ and most sets of states $\rho_i$, however, the observables $\Lambda(\rho_i)$ have entangled eigenstates, and thus this expectation value is impractical to measure experimentally. It is thus necessary to express the process fidelity in terms of a complete set of mutually orthonormal, local observables $W_j$ as
\be
  F(\Lambda, \Gamma) = \frac{1}{d^2} \sum_{ij=1}^{d^2} C_{ij}\tr[\Gamma(\rho_i)W_j]\ ,
  \label{eq:process_fidelity_expanded}
\ee
with
\be
C_{ij}=\sum_{l=1}^{d^2} [B^{-1}]_{li} \tr[\Lambda(\rho_l)W_j]\ .
\ee
The full experimental protocol entailed by Eq.~\eqref{eq:process_fidelity_expanded} implies the preparation of $d^2$ initial states $\rho_i$ and the measurement of $d^2$ observables $W_j$ per initial state for a total of $d^4$ experimental settings.
Since the dimension $d$ grows exponentially in the number of qubits, the experimental effort required to evaluate the process fidelity is prohibitively high even for a moderate number of qubits.

It is, however, possible to {\em estimate} the process fidelity using far fewer experimental settings~\cite{flammia2011direct,da2011practical}.
The procedure involves sampling a small subset of input states and measurement bases from a joint probability distribution which guarantees the resulting estimates will have a low variance regardless of the specific channels being evaluated.

Unfortunately, this protocol may only be applied to the process fidelity as expressed in Eq.~\eqref{eq:process_fidelity}. The form of Eq.~\eqref{eq:process_fidelity_expanded} precludes the definition of an estimator with similarly favourable statistical properties. The estimation protocol thus implicitly relies on the inputs being orthonormal, meaning that not all of them can be quantum states. An experimental protocol may still be obtained by sampling quantum input states from the eigenbasis of each $\sigma_i$ input on a shot-by-shot basis, however this relies on the ability to vary the experimental setup at every shot of the experiment.

Given the operation speed of current laboratory control software this can reduce the repetition rate of an experiment substantially. Consistently with this, the interfaces to currently available NISQ devices limit the number of initial states and measurement settings that can be explored, whereas they do not impose comparably severe limitations to the repetition of the same experiment \textit{i.e.} with the same initial state and measurement basis~\cite{Qiskit2}.

In NMR quantum computing~\cite{jones2010quantum,lu2016nmr,vandersypen2005nmr,somaroo1999quantum}, the situation is even more drastic, since expectation values are obtained from the simultaneous measurement of an ensemble of qubits rather than through individual projective measurements~\cite{cory1997ensemble}. Varying the initial states and observables thus necessarily increases the overhead by a factor of the number of chosen experimental settings, severely hindering the efficiency of the protocol.

The goal of this work is to develop an alternative formulation for assessing the quality of implemented quantum channels based on Eq.~\eqref{eq:process_fidelity_expanded} that may be efficiently estimated without resorting to frequent changes in initial state preparation and final measurement basis.

\subsection{The hierarchy of $\mathbf{k-}$fidelities}\label{sec:k_fidelity}

In order to use the process fidelity as in Eq.~\eqref{eq:process_fidelity_expanded}, a set of states which span the space of linear operators must be specified. The natural choice is to take these states to be as close to orthogonal as possible, which may be achieved by minimising $\sum_{i\neq j}\tr[\rho_i \rho_j]$. For a single qubit, the analytical solution to this is any set of four states which form the vertices of a regular tetrahedron centred at the origin of the Bloch sphere. These states are known as {\it symmetric informationally complete} (SIC) states~\cite{renes2004symmetric}.

A set of SIC states can also be defined for the full multi-qubit system, but it would typically contain entangled states.
In order to keep state preparation errors to a minimum, it is desirable to have only separable initial states. As shown in Appendix~\ref{ap:optimal_states}, the set of states formed by taking the $n-$fold tensor product of the single qubit SIC states minimises $\sum_{i\neq j}\tr[\rho_i \rho_j]$ among all complete sets of product states, and we therefore take this set as initial states in the following.
The inverse of the overlap matrix $B_{ij}$ is then readily obtained. The inverse of the overlap matrix $B^{(1)}$ for a single qubit reads $\left[B^{(1)}\right]^{-1} = \mathds{1}_4 - A$, in terms of the matrix $A$ with elements $A_{ij} = \frac{1}{4}(1-5\delta_{ij})$. In the case of $n$ qubits, the inverse of $B$ is the $n$-fold tensor product $(\mathds{1}_4 - A)^{\otimes n}$. Collecting terms with a given number of factors of $A$, this is expressed as
\begin{equation}\label{eq:B_expansion}
   B^{-1} = \sum_{k=0}^n (-1)^k \Omega_k \ ,
\end{equation}
with
\be
\Omega_k=\sum_\pi \underbrace{\mathds{1}_4\otimes\mathds{1}_4\otimes\hdots\otimes\mathds{1}_4}_{k}\otimes\underbrace{A\otimes A\otimes\hdots\otimes A}_{n-k}\ ,
\ee
where the sum is performed over all inequivalent permutations of identity and operators $A$.

By truncating Eq.~\eqref{eq:B_expansion} at different values of $k$, one can define a hierarchical series of \textit{$k-$fidelities}. Truncating the expansion at the $k-$th term means that only pairs of input states differing by at most $k$ single qubit states will have non-zero contributions to the $k-$fidelity. The highest order term, with $k=n$, retains all orders within the sum (Eq.~\eqref{eq:B_expansion}) and is thus not an approximation, but it coincides exactly with the process fidelity.

The lower the overlap between $\rho_i$ and $\rho_j$, the less that pair of states contributes to the overall $k-$fidelity. The coefficient
\be
    c_m = \sum_{j=m}^n (-1)^{2j + m} \left( \frac{1}{4} \right)^j {n-m \choose j-m}
    = (-1)^m \frac{5^{n-m}}{4^n}
    \label{eq:order_coefficient}
\ee
quantifying this contribution for a pair of states differing by $m$ single qubit states decreases exponentially in $m$. Thus the higher the order $k$ in Eq.~\eqref{eq:B_expansion}, the lower the corresponding coefficient in the $k-$fidelity. The leading order term, the {\em $0-$fidelity},
\begin{equation}
  F_0(\Lambda,\Gamma) = \frac{1}{d^2} \sum_{ij=1}^{d^2} \tr[ \Lambda(\rho_i)W_j]\tr[\Gamma(\rho_i)W_j] \ ,
  \label{eq:0_fidelity}
\end{equation}
may then be used as an approximation to the process fidelity with highly convenient properties which will be shown in the following section. Moreover, the $0-$fidelity may be efficiently estimated using a small number of unique experiments, without requiring expensive shot-by-shot changes to the input state and measurement basis.

\subsection{Properties of the $\mathbf{0-}$Fidelity}\label{sec:0_fidelity_properties}
There are a number of properties that the $0-$fidelity satisfies which make it an effective proxy for the process fidelity:

\begin{itemize}
\item[(i)] \textit{Faithfulness}:
The $0-$fidelity is maximised if and only if the process fidelity is also maximised, \textit{i.e.} if the two channels being considered are identical.
This means that the $0-$fidelity is faithful~\cite{cerezo2021cost}
\item[(ii)] \textit{Monotonicity}: It is an approximately monotonic function of the process fidelity, meaning that high fidelity channels give rise to high $0-$fidelities, with only small deviations from monotonicity which decrease at high fidelities.
\item[(iii)] \textit{Scalability}: Finally, as the system size increases, the $0-$fidelity becomes an increasingly better approximation to the process fidelity.
\end{itemize}

In the following section these properties will be shown through analytical proof for (i) and through numerical evidence for (ii) and (iii).

(i) {\it {Proof of faithfulness}}
Following Eq.~\eqref{eq:general_process_fidelity} the $0-$fidelity  can be expressed as
\be
    F_0 = \frac{1}{d^2}\sum_{i=1}^{d^2}\tr[\Lambda(\rho_i)\Gamma(\rho_i)] \ ,
\ee
which can be understood as a sum over state fidelities between the states $\Lambda(\rho_i)$ and $\Gamma(\rho_i)$.
For unitary target channels $\Lambda$ and pure input states $\rho_i$, each state fidelity is maximised if and only if the states are identical, thus the $0-$fidelity is maximised if and only if
\be
  \Gamma(\rho_i) = \Lambda(\rho_i) \ \ \forall \rho_i \ .
\ee
Since the input states form an operator basis by construction, any arbitrary operator $\mathcal{O}$ may be written as a linear combination ${\mathcal{O} = \sum_i c_i \rho_i}$. It therefore follows that ${\sum_i c_i \Lambda(\rho_i) = \sum_i c_i \Gamma(\rho_i)}$ which implies (through the linearity of quantum channels) that for any $\mathcal{O}$, ${\Lambda(\mathcal{O}) = \Gamma(\mathcal{O})}$. The $0-$fidelity is therefore maximised if and only if $\Lambda = \Gamma$ and since this is also the necessary condition for the process fidelity to be maximised it therefore follows that the $0-$fidelity is faithful.

\begin{figure}
  \centering
   \includegraphics{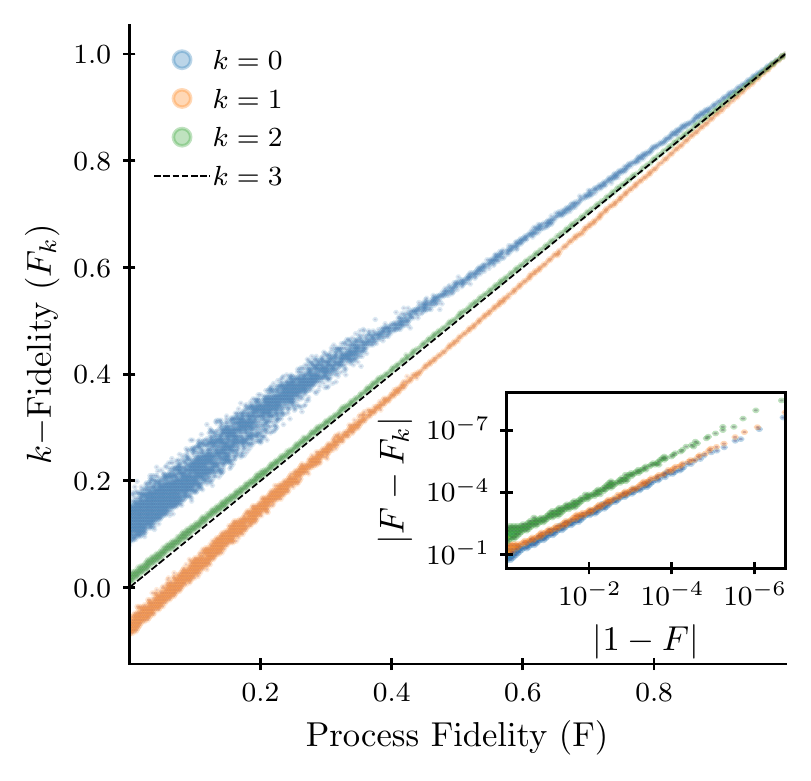}
   \caption{Plot of $k-$fidelity against process fidelity (evaluated numerically using Eq.~\eqref{eq:process_fidelity}) for $10000$ numerical evaluations of randomly generated 3 qubit unitary channels compared with a random unitary target. As the order $k$ increases, the $k-$fidelity more closely approximates the process fidelity, with the $3-$fidelity being equal to it. At high fidelities, all orders give rise to approximately monotonic, close approximations to the process fidelity, with the deviations becoming very small at very high fidelities as shown in the inset.}
   \label{fig:B_expansion}
\end{figure}

(ii) {\it {Monotonicity}}
Fig.~\ref{fig:B_expansion} shows how well each order of the $k-$fidelity approximates the process fidelity in the assessment of three qubit quantum channels, generated numerically by comparing $10 000$ randomly generated unitary channels with a fixed random unitary target channel. All the orders converge to the process fidelity at high values, as can be seen in the inset, with the $3-$fidelity corresponding precisely to the process fidelity evaluated using Eq.~\eqref{eq:process_fidelity}. The process by which the random unitaries are generated is outlined in the Appendix.

For orders $k<n$, the $k-$fidelities are not true monotonic functions of the process fidelity, however they are {\em approximately} monotonic (this may be seen in the thickness of the evaluated points). The $0-$fidelity, as the lowest order, deviates most substantially from monotonicity. Nevertheless it is close to monotonic for fidelities above $\sim50\%$ and even at low fidelities the deviations from monotonicity are relatively small, meaning that the $0-$fidelity should still be useful as an approximation to the process fidelity for quantum channel optimisation.

Unlike the process fidelity, for some orders $k$ the corresponding $k-$fidelity may have negative values as seen in the negative $1-$fidelities at very low fidelities in Fig.~\ref{fig:B_expansion}. This negativity arises from the negative coefficient for $m=1$ in Eq.~\eqref{eq:order_coefficient}; for most fidelities this is counterbalanced by the fact that the $k-$fidelity is dominated by the zeroth order terms (which are non-negative by construction) however at low fidelities these can be exceeded by the negative terms at higher orders.

\begin{figure}
  \centering
   \includegraphics[width=\linewidth]{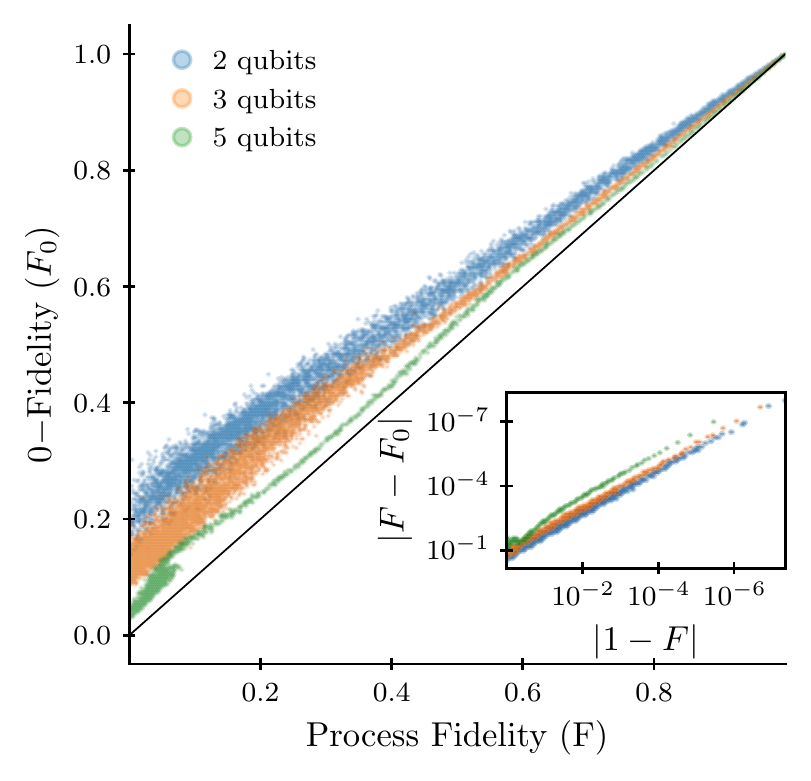}
   \caption{Plot of $0-$fidelity against process fidelity for $10000$ numerical evaluations of randomly generated 2, 3 and 5 qubit unitary channels evaluated against random unitary targets. The $0-$fidelity approximates the process fidelity increasingly well as the number of qubits increases. The inset shows that at high fidelities the $0-$fidelity converges to the process fidelity for all system sizes.}
   \label{fig:FOM_comparison}
\end{figure}

(iii) {\it {Scalability}}
As the number of qubits increases the process fidelity is increasingly well approximated by the $0-$fidelity (Fig.~\ref{fig:FOM_comparison}). The $n=5$ qubit case (green) consistently gives rise to $0-$fidelity values which are close to the process fidelities, and additionally exhibits the desired monotonicity above $\sim 15\%$ whilst the $n=2$ qubit case (blue) only exhibits this above $\sim 90\%$. This strongly suggests that the $0-$fidelity may be relied upon as a process fidelity approximation even for relatively large systems.

\section{Estimation of the $0-$ and Process Fidelities}
\subsection{Estimating the $\mathbf{0-}$Fidelity}\label{sec:estimation_fom}
As with the process fidelity, evaluating the $0-$fidelity involves a number of experimental settings scaling exponentially with the system size and thus for a practical experimental prescription it is necessary to estimate it using random sampling~\cite{flammia2011direct,da2011practical}. The protocol works by choosing input states and measurement bases from the joint probability distribution
\be
    \operatorname{Pr}(i,j) = \frac{1}{d^2}\tr[\Lambda(\rho_i)W_j]^2\ .
    \label{eq:prob_dist}
\ee
These selected settings may then be used to construct an estimator
\begin{equation}
  X(i,j) = \frac{\trace[\Gamma(\rho_i)W_{j}]} {\trace[\Lambda(\rho_i)W_{j}]}\ ,
  \label{eq:estimator}
\end{equation}
in terms of experimentally measurable quantities $\trace[\Gamma(\rho_i)W_{j}]$; the expectation value of $X$ over $i$ and $j$ is equal to the $0-$fidelity.

The crucial factor which makes this procedure extremely useful lies in the sample variance of the estimator,
\be
    \operatorname{Var}(X) = 1 - F_0^2 \ ,
    \label{eq:variance}
\ee
which is bounded by $1$ independently of the specific channels being assessed. This means that by taking the mean of $l$ evaluations of $X(i,j)$ sampled from Eq.~\eqref{eq:prob_dist} an unbiased estimate of the $0-$fidelity may be obtained with a sample variance of, at worst, $1/l$. Manageable sample variances may be achieved by evaluating a relatively small ($l\sim 100$) number of expectation values, a number which does not scale with the system size.

For the variance specified in Eq.~\eqref{eq:variance} it is implicitly assumed that the terms $X(i,j)$ given in Eq.~\eqref{eq:estimator}, are obtained exactly, however the accuracy of these terms is limited by experimental constraints. In platforms such as superconducting qubits, the accuracy of estimating $X(i,j)$ is limited by the number of experimental repetitions (shots) $m$ required to acquire expectation values from projective measurements, a number which depends on the details of the channel $\Lambda$ and which scales in the worst case as $O(d)$ (bounds on the variance of the $0-$fidelity estimates are given in Appendix~\ref{ap:variance_bounds}). In NMR platforms, where expectation values are obtained from a single experiment, the primary limiting factor is the finite measurement acquisition time which limits the resolution at which expectation values may be extracted.

The two distinct platforms entail slightly different implementations. In both cases, $l$ experimental settings $(i,j)$ corresponding to expectation values $\tr[\Gamma(\rho_i)W_j]$ are selected according to the probability distribution Eq~\eqref{eq:prob_dist}. In {\em projective estimation}, the expectation values are then obtained experimentally by running each setting $m$ times and taking the average of the projective measurements, whilst in {\em full trace estimation}, the expectation values are obtained exactly from a single experimental measurement.

For estimations of the $0-$fidelity the distinction between the two implementations is merely a technical detail, however for estimating the process fidelity the choice of experimental platform can have a severe impact on the quality of the estimations.

\subsection{Estimating the Process Fidelity}

In order to compare the quality of the estimation protocol for the $0-$fidelity to the equivalent protocol for estimating the process fidelity, it is necessary to expand Eq.~\eqref{eq:process_fidelity} in a local orthonormal basis to obtain an expression which is analogous to Eq.~\eqref{eq:0_fidelity} as
\be
    F(\Lambda, \Gamma) = \frac{1}{d^2}\sum_{ij=1}^{d^2}\tr[\Lambda(\sigma_i^\dagger)W_j] \tr[\Gamma(\sigma_i)W_j]\ .
    \label{eq:process_fidelity_flammia}
\ee
Since the input operators in Eq.~\eqref{eq:process_fidelity_flammia} are not quantum states, an additional step is required in order to obtain experimentally realisable settings. Once a measurement setting corresponding to an expectation value $\tr[\Gamma(\sigma_i)W_j]$ has been chosen, an experimental prescription may be obtained by expanding $\sigma_i$ in its eigenbasis.
This results in the relation
\be
    \tr[\Gamma(\sigma_i)W_j] = \sum_{k=1}^d \lambda^{\sigma_i}_k \tr[\Gamma(|\phi^{\sigma_i}_k\rangle \langle \phi^{\sigma_i}_k|)W_j]\ ,
    \label{eq:eigenbasis_expansion}
\ee
where $|\phi^{\sigma_i}_k\rangle$ are eigenstates of $\sigma_i$ with corresponding eigenvalues $\lambda^{\sigma_i}_k$. An appropriate choice of operators $\sigma_i$ (for example, the set of tensor products of normalised Pauli operators) means these states will be separable, and thus can be implemented experimentally. This implies an increase in the number of expectation values which need to be experimentally evaluated by a factor of $d$.

In projective estimations it is possible to obtain estimates of the process fidelity using the same total number of experiments $lm$ as that required for $0-$fidelity estimation by sampling experimental input states from the eigenbasis of $\sigma_i$ on a shot-by-shot basis~\cite{flammia2011direct,da2011practical}; as discussed in Sec.~\ref{sec:quality_channels} however, this results in inefficiencies which render the strategy inapplicable on NISQ devices.

Nevertheless, projective estimations of the process fidelity may still be obtained using the same number of expectation values $l$ and the same total number of experiments $lm$ as the $0-$fidelity. For any expectation value $\tr[\Gamma(\sigma_i)W_j]$, implementing all $d$ eigenstate expectation values $\tr[\Gamma(|\phi^{\sigma_i}_k\rangle \langle \phi^{\sigma_i}_k|)W_j]$ with each experiment repeated $m/d$ times is equivalent to estimating $\tr[\Gamma(\sigma_i)W_j]$ using $m$ shots. Thus the total number of experiments remains the same between the process and $0-$fidelity estimations, with the caveat that estimating the process fidelity requires $dl$ {\em unique} experiments as compared with only $l$ for the $0-$fidelity. The bounds on the variance of the process fidelity estimates are given in Appendix~\ref{ap:variance_bounds}.

In the case of full trace estimation, each unique experiment is only repeated once, thus the total number of experiments is $l$. In this case, implementing all $d$ eigenstates of each $\sigma_i$ necessarily entails a factor of $d$ increase in experimental overhead over estimating the $0-$fidelity.

\subsection{Comparing the Process and $\mathbf{0-}$Fidelity Estimates}\label{sec:estimation_comparisons}
In this section the quality of process and $0-$fidelity estimations are compared in both the full trace and projective experimental situations. In line with current implementations of NISQ devices, in the following numerical simulations the maximum number of unique experiments is limited to a maximum value of $900$. Details on the generation of random channels can be found in the Appendix.

Fig.~\ref{fig:f_vs_FOM_dists} shows deviations of the full trace estimates of the process (blue histogram) and $0-$fidelities (green histogram) from their true values,
based on a randomly generated three qubit unitary target channel and a perturbed test channel. The estimate of the $0-$fidelity is realized in terms of $l=160$ settings $\tr[\Gamma(\rho_i)W_j]$, with each expectation value evaluated exactly. The process fidelity is sampled with $l=20$ settings $\tr[\Gamma(\sigma_i)W_j]$ for all $d=8$ eigenstates $\tr[\Gamma(|\phi^{\sigma_i}_k\rangle \langle \phi^{\sigma_i}_k|)W_j]$ evaluated exactly per setting, for a total of $dl=160$ experimental settings.

Fig.~\ref{fig:f_vs_FOM_dists} exemplifies the fact that the statistical fluctuations in estimates of the $0-$fidelity are much smaller than those of the process fidelity, with the standard deviation of the $0-$fidelity estimation errors being $\sim 0.05$ compared to $\sim 0.16$ for the process fidelity.

\begin{figure}
  \includegraphics[width=\linewidth]{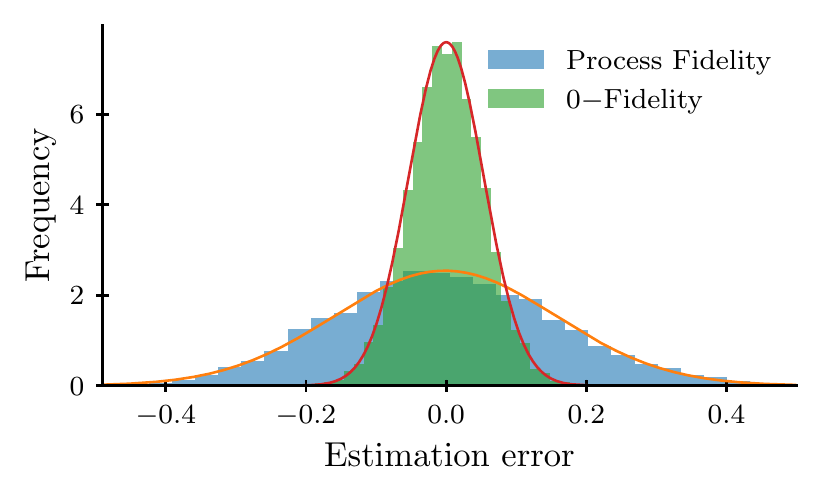}
  \caption{Statistical deviations from their true values for $10000$ numerically evaluated full trace estimations of the process and $0-$fidelities, each using $l=160$ expectation values per evaluation. The estimations were made using the overlap between a random three qubit target unitary and a perturbed test unitary, with the same setup for both the process and $0-$fidelity estimations. The variance of the $0-$fidelity estimations is significantly lower than those of the process fidelity, making it far more suitable for an optimisation cost function.}
  \label{fig:f_vs_FOM_dists}
\end{figure}

For the case of projective estimations, the number of shots $m$ may be adjusted such that the total number of experiments required for estimating the process fidelity is the same as that required to estimate the $0-$fidelity. Even in this favourable setting for the process fidelity, it is still substantially outperformed by the $0-$fidelity estimates.

\begin{figure}
  \includegraphics[width=\linewidth]{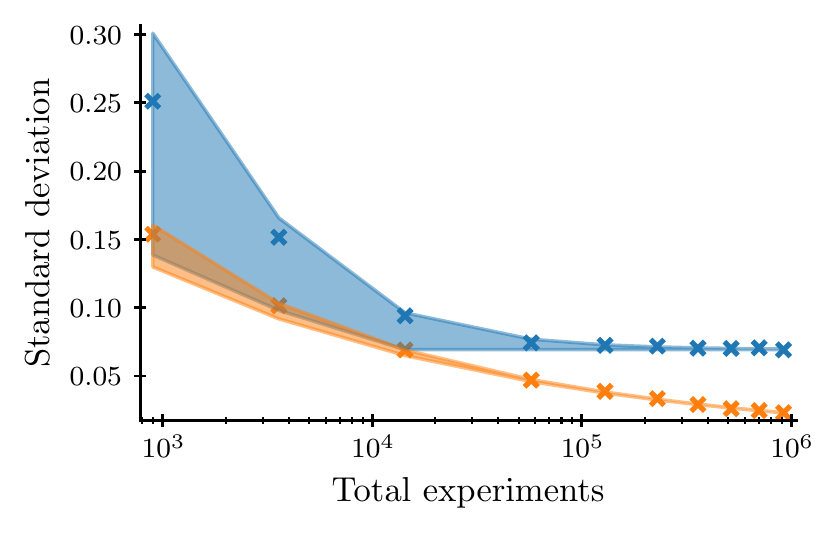}
  \caption{Plot of the standard deviation of process and $0-$fidelity projective estimations as the number of experiments increases. The data are generated numerically through $10000$ estimations per data point of the comparison between a pair three random unitary channels, with the same pair used to generate all data points and where the number of expectation values $l$ was approximately equal to the number of shots $m$. The $0-$fidelity standard deviations are consistently below those of the process fidelity estimates regardless of the total number of experiments used. Moreover, the process fidelity standard deviations seem to saturate to a minimum value, whereas no such saturation can be observed for the $0-$fidelity. The shaded regions correspond to the upper and lower bounds on the estimations using the analytical bounds in Appendix~\ref{ap:variance_bounds}.}
  \label{fig:std_progression}
\end{figure}

Fig.~\ref{fig:std_progression} shows the standard deviations of the estimation error for $10000$ estimations of the process and $0-$fidelities of a randomly generated pair of three qubit unitary channels (with the same pair being used for all data points) as the total number of experiments $lm$ increases. The details for the specific generation of the random channels, along with the experimental parameters $l$ and $m$ used are outlined in the Appendix. The $0-$fidelity estimates have much lower standard deviations than those of the process fidelity at every allocation of (numerically generated) experiments investigated.

A choice of $l\approx m$ expectation values and measurement settings gives rise to estimates with the lowest variance for a given total number of experiments $lm$. Since for the process fidelity the measurement of $l$ expectation values implies the implementation of $dl$ unique experiments, the maximum number of expectation values which can be measured is limited to $l \leq \lfloor 900/d \rfloor$, meaning that if one wants to use a relatively high number of experiments in order to obtain an estimate, it is necessary to use a suboptimal allocation of experiments $l<m$. This is reflected in Fig.~\ref{fig:std_progression}, which shows that the standard deviations for the process fidelity seem to saturate to a minimum value once the maximum number of unique experiments is reached. An equivalent limit will eventually be reached for the $0-$fidelity, however this limit is higher than that constraining the process fidelity by a factor of $d$ and, moreover, is independent on the size of the channel being evaluated.

Although the limitation to 900 unique experiments is in some sense arbitrary (being imposed by the provider of the quantum hardware) it reflects the fact that implementing many different experiments is more expensive than repeating a single experiment many times; were this restriction to be lifted one would expect the standard deviation to no longer saturate. Even in this case, for any given number of experiments the $0-$fidelity estimates will have a lower standard deviation than the process fidelity estimates as shown by the bounds given in Appendix~\ref{ap:variance_bounds}.

\begin{figure}
  \includegraphics{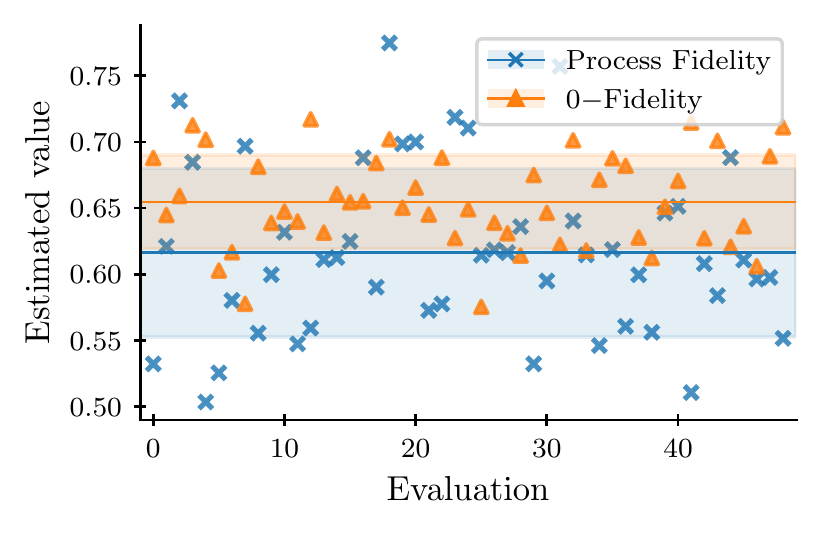}
  \caption{Plot showing 50 estimations of the process and $0-$fidelities for a single random three qubit target unitary implemented on the \texttt{ibmq\_toronto} quantum device. The mean for the $0-$fidelity (orange line) is slightly higher than that of the process fidelity (blue line), since the $0-$fidelity overestimates the process fidelity as shown in Fig.~\ref{fig:FOM_comparison}, however the standard deviation of the $0-$fidelity (orange shaded region) is substantially lower than that of the process fidelity (blue shaded region).}
  \label{fig:toronto_statistics_evals}
\end{figure}

The superior performance of the $0-$fidelity estimates is also reflected in real experimental data. Fig.~\ref{fig:toronto_statistics_evals} shows $50$ estimations of the process (blue crosses) and $0-$fidelities (orange triangles) for a random three qubit quantum circuit implemented on the \texttt{ibmq\_toronto} quantum computer (where the circuit would perfectly implement the target channel in the absence of noise). The random circuit used is given in Appendix~\ref{ap:rand_channels}.

The $0-$fidelity estimations were performed using $l=336$ circuits (that is, $336$ unique experimental settings) each using $m=336$ shots, which may be evaluated efficiently enough to permit an optimisation involving $\sim150$ iterations to be performed whilst still resulting in relatively low variances. The settings for the process fidelity estimations were then found by taking the maximum possible number of unique settings ($dl=896$ circuits, corresponding to $l=112$ expectation value evaluations $\tr[\Gamma(\sigma_i)W_k]$) and setting the number of shots to $m=144$ such that the total number of experiments was equal to that used for the $0-$fidelity estimations.

The $0-$fidelity data are clustered much closer together than the process fidelity, indicating that estimates of the $0-$fidelity are more suitable as an efficient protocol for evaluating the quality of quantum channels. Although it is not feasible to experimentally perform enough estimates to fully capture the statistics for the distributions as in the numerical analysis above, the data clearly indicate that the $0-$fidelity estimations have a much lower standard deviation than the process fidelity estimates, with the standard deviation of the $0-$fidelity estimations being $\sim0.033$ as compared with $\sim0.07$ for the process fidelity. These are in line with what one would expect from the analytical bounds derived in Appendix~\ref{ap:variance_bounds}, which yield upper bounds of $0.09$ and $0.05$ for the process and $0-$fidelity respectively.

\section{Gate Optimisation}\label{sec:optimisation}
Estimates of the $0-$fidelity are a highly efficient way of evaluating the quality of noisy quantum channels. One application for this is in the variational optimisation of such channels, in which the parameters of a parameterised channel are varied according to some classical optimisation algorithm until the $0-$fidelity is maximised.

In the following section the results of optimisations performed using Bayesian optimisation (BO)~\cite{movckus1975bayesian} are presented. BO is highly efficient and resilient to noise, as demonstrated in its successful application in related problems in quantum optimal control~\cite{sauvage2020optimal,mukherjee2020preparation,mukherjee2020bayesian,vqa_sharing,tham2016simulating,antonik2020bayesian,craigie2020resource}. For the interested reader, a thorough review of BO may be found in Refs~\cite{frazier2018tutorial,shahriari2015taking,williams2006gaussian}.

\begin{figure}
    \includegraphics{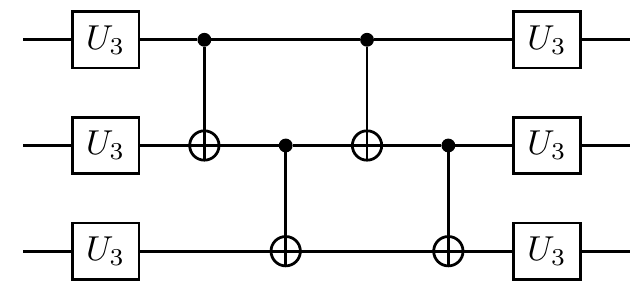}
  \caption{Circuit diagram for the implementation of a parameterised CNOT gate between the first and third qubits, which are assumed to not be physically connected in the device. Each of the $U_3$ gates has three parameters over which the optimisation may be performed - the ideal gate may be implemented on a noise-free device by setting all the parameters to 0.
  }
  \label{fig:3q_cnot}
\end{figure}

The target channel in these optimisations was a CNOT gate between non-connected qubits, a three qubit channel necessitated by the limited connectivity of NISQ devices~\cite{murali2019full}. The ideal channel may be implemented using only CNOT gates, however a parameterised version may be generated by appending and prepending single qubit gates on all qubits as seen in Fig.~\ref{fig:3q_cnot}.

It should be noted that in the following the effects of state preparation and measurement error are not taken into account. These effects may be addressed using, for example, self-consistent tomography~\cite{merkel2013self} or techniques developed in randomised benchmarking~\cite{carignan2015characterizing,helsen2020general,flammia2020efficient,harper2020efficient} however such techniques are not considered in this work.

\subsection{Optimisation Results}
\begin{figure}
  \includegraphics{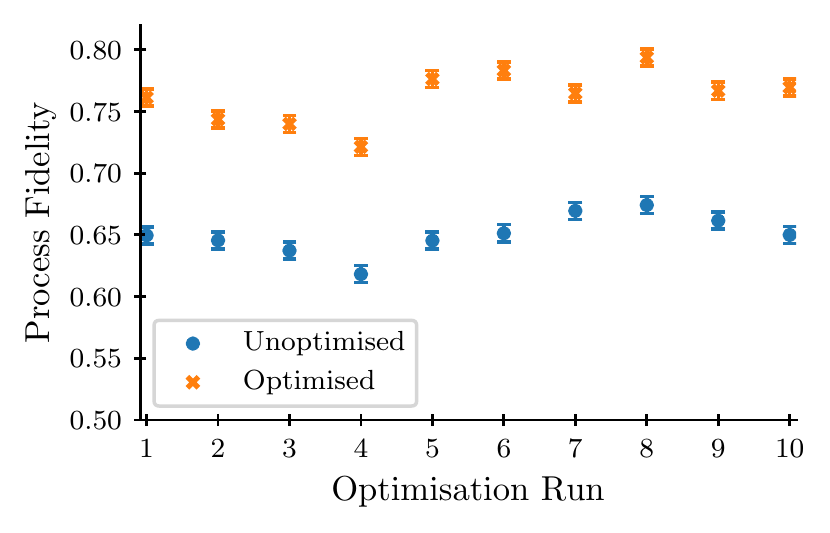}
  \caption{Plot showing the improvement obtained by the Bayesian optimisation of the circuit in Fig.~\ref{fig:3q_cnot} for 10 optimisation runs given in terms of the process fidelity evaluated using all $d^4$ measurement settings, implemented on the \texttt{ibmq\_singapore} quantum device. Each optimisation run was performed using 140 iterations using $0-$fidelity estimations obtained using 160 unique circuits, each repeated for 2048 shots, with the final results evaluated using 4096 circuits, each using 8192 shots, with the error bars corresponding to the $95\%$ confidence interval arising from the finite number of shots. The optimised results consistently outperform the unoptimised results, demonstrating the effectiveness of using the $0-$fidelity as the figure of merit for quantum channel optimisation.}
  \label{fig:optimised_res}
\end{figure}

The results of the optimised circuits based on Fig.~\ref{fig:3q_cnot} are shown in Fig.~\ref{fig:optimised_res} in terms of the process fidelity evaluated using all $d^4$ measurement settings in Eq.~\eqref{eq:process_fidelity_expanded}. The actual optimisations were performed using estimates of the $0-$fidelity however evaluating the final channels through the fully evaluated process fidelity allows for direct comparisons with previous work. This also provides further evidence that the $0-$fidelity is an excellent proxy for the process fidelity in this setting.

The optimised circuits (orange triangles) consistently achieved substantially higher fidelities than their unoptimised counterparts (blue circles) over every experimental run. The average process fidelity for the unoptimised runs was $0.65$, whilst the optimisation yielded circuits with average fidelities of $0.76$, a significant ($\sim 17\%$) relative improvement. The error bars correspond to the $95\%$ confidence interval arising from the statistical variation due to shot noise.  The variation in process fidelity between optimisation runs is a consequence of the fact that each run was performed on a different day, over which time the properties of the device change due to drifts and recalibration, giving rise to slightly higher or lower fidelities; these fluctuations are significantly larger than the uncertainties in the process fidelity measurements and so it is unlikely that these effects arise from finite sampling effects.

\begin{figure}
  \includegraphics{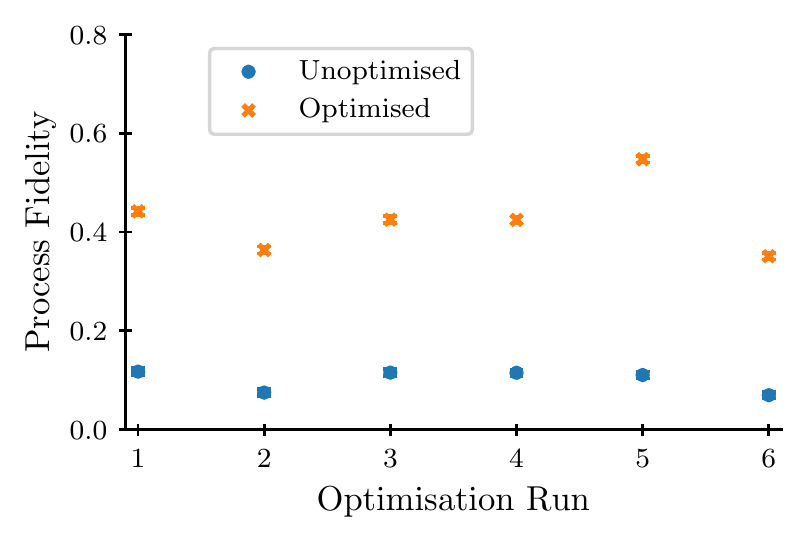}
  \caption{Plot showing the process fidelity (evaluated using all $d^4$ measurement settings) when the optimised and unoptimised channels are applied three times (equivalent in the absence of noise to a single application) for 6 runs of the Bayesian optimisation of the circuit in Fig.~\ref{fig:3q_cnot} implemented on the \texttt{ibmq\_singapore} quantum device. Each optimisation run was performed using 140 iterations using $0-$fidelity estimations obtained using 160 unique circuits, each repeated for 2048 shots, with the final results evaluated using 4096 circuits, each using 8192 shots, with the error bars corresponding to the $95\%$ confidence interval arising from the finite number of shots. The optimised results are substantially better across all experimental runs, with the optimisation increasing the average process fidelity from $0.11$ to $0.44$.}
  \label{fig:optimised_res_meas_err}
\end{figure}

The ultimate goal of optimising quantum channels is to use them as part of a larger algorithm. As such, it is critical that this increase in fidelity is retained when such a composition is performed. The results shown in Fig.~\ref{fig:optimised_res_meas_err} confirm that this is the case: here the experimental channel was the CNOT applied three times, which should be equivalent to a single CNOT in the absence of noise. Once again, the optimised circuits attain higher fidelities in every experimental run, substantially outperforming the textbook implementations with average fidelities increasing from $0.11$ to $0.44$ (the error bars reported in the figure again correspond to the $95\%$ confidence interval associated with shot noise).

We attribute these gains to the ability of the BO to find parameters which counteract gate errors in the implemented circuit, which it is able to do without any formal characterisation of the form of those errors. If the optimisation were merely counteracting state preparation and measurement errors, one would not expect the gains in fidelity to be maintained when the circuit is applied multiple times.

Moreover, the gains observed for the repeated application of the optimised gate are greater than for a single application. If the gate errors were entirely stochastic, applying the gate three times would result in an overall error which is approximately the cube of the individual gate error. This is observed in the optimised results ($0.76^3 \approx 0.44$) however for the unoptimised results the error from applying the gate three times is substantially higher than expected. This may be attributed to \textit{systematic} errors which combine constructively with multiple applications, resulting in overall errors which are larger than if the errors were independent and stochastic. The fact that this effect disappears in the optimised case suggests that the BO was able to eliminate the bulk of the systematic gate error. The advantageous properties of stochastic errors are well known, forming the basis for the technique of \textit{randomised compiling}~\cite{wallman2016noise} in which additional gates are applied in order to convert coherent noise sources into stochastic ones. The fact that such a strategy has practical advantages demonstrates the utility of the optimisation algorithm, since here the systematic errors are not only converted but are directly reduced.

\section{Conclusions}
NISQ devices, whilst having much potential are limited by their inherent noisiness. Quantum channels may be directly optimised such that the resulting channel produces a much more faithful implementation of the desired dynamics. In order to perform such an optimisation, a figure of merit is required which can efficiently characterise the quality of an implemented channel. Direct process fidelity estimation would be a natural choice for this, however its implementation on NISQ hardware is rendered impractical through the requirement that the input state and measurement basis be changed at every shot of an experiment.

In this work we present an alternative figure of merit which overcomes this issue. The leading order term in a hierarchical series of $k-$fidelities, the $0-$fidelity, is a faithful approximation to the process fidelity which can be estimated on any quantum platform using only a small number of unique experiments that does not scale with system size. Estimates of the $0-$fidelity substantially outperform estimates of the process fidelity under the constraints imposed by current implementations of NISQ devices, with this advantage being demonstrated both numerically and experimentally.

The $0-$fidelity is an excellent figure of merit for the direct optimisation of quantum channels. This is demonstrated through the successful optimisation of a CNOT channel on an IBM Quantum device, for which we report significant ($\sim17\%$) improvements over the textbook implementation.

\section{Outlook}
We envisage our $0-$fidelity-based optimisation routine could be applied to concrete problems in quantum simulation. One potential route could involve splitting an algorithm up into small blocks, optimising each one and then composing the optimised channels such that the overall simulation is less noisy. This is particularly relevant to simulations of Trotterised quantum dynamics and algorithms such as the quantum approximate optimisation algorithm (QAOA), where layers of short-depth quantum circuits are repeated many times.

Additionally, it would be instructive to implement the protocol on other NISQ platforms. Whilst the framework presented here has been given in terms of digital quantum gates, it is equally applicable to low level pulse-based quantum control and should be implementable on any quantum platform. Implementations on an NMR quantum computer would be of particular interest as the advantages offered by the $0-$fidelity over the process fidelity are more substantial.

On a more theoretical basis, it would be interesting to investigate whether similar hierarchical structures can be derived when the restriction to product states is relaxed. This would likely have a significant impact on the practical implementation of the protocol, but may yield some interesting insight as to the structure of SIC states. An additional route for further work would be to combine the $0-$fidelity with techniques for alleviating SPAM errors such as cycle benchmarking~\cite{erhard2019characterizing} and randomised benchmarking~\cite{carignan2015characterizing,helsen2020general,harper2020efficient}. Finally, it is known that for Pauli channels the process fidelity can be efficiently evaluated by measuring the eigenvalues of the superoperator matrix~\cite{nambu2005matrix} associated with the quantum channel~\cite{flammia2020efficient}. A potentially valuable route for further work would be to investigate any potential relationship between the hierarchy of $k-$fidelities and this previous work, particularly since the $k-$fidelities are not restricted to Pauli channels.

\section{Acknowledgments}
We are grateful to Rick Mukherjee for providing stimulating discussions. This work is supported by Samsung GRP grant, the UK Hub in Quantum Computing and Simulation, part of the UK National Quantum Technologies Programme with funding from UKRI EPSRC grant EP/T001062/1 and the QuantERA ERA-NET Co-fund in Quantum Technologies implemented within the European Union’s Horizon 2020 Programme. S.G. and F.S. are supported by studentships in the Quantum Systems Engineering Skills and Training Hub at Imperial College London funded by EPSRC (EP/P510257/1). We acknowledge the use of IBM Quantum services for this work. The views expressed are those of the authors, and do not reflect the official policy or position of IBM or the IBM Quantum team. Numerical simulations were carried out on Imperial HPC facilities~\cite{imperialHPC}.

\section{Data Availability}
All of the data presented in this paper, along with the python code used to generate this data and the figures, can be found in a public GitHub repository~\cite{codeRepo}.

\appendix
\section{Bounds on the variance of the fidelity estimations}\label{ap:variance_bounds}
\subsection{$0-$Fidelity Estimates}
Here we derive bounds on the variance of an estimate of the $0-$fidelity obtained using $l$ unique experimental settings corresponding to the terms $\tr[\Gamma(\rho_i)W_j]$ which are estimated as the mean of $m$ projective measurements in the eigenbasis of $W_j$. To proceed, we take $\{W_j\}$ to be the set of normalised Pauli operators, meaning that each $W_j$ has eigenvalues $\lambda^{W_j}_k \in \{\pm 1/\sqrt{d}\}$.

It follows from the orthonormality of $\{W_j\}$ and from the fact that $\tr(\rho^2)\leq 1$ for any arbitrary state $\rho$ (with equality holding if $\rho$ is pure) that
\begin{equation}
\sum_j \tr[\mathcal{E}(\rho_i)W_j]^2 \leq 1,
\end{equation}
for an arbitrary quantum channel $\mathcal{E}$. It thus follows that the variance $\Delta^2_{(i,j)}$ of a single measurement corresponding to a single setting $\tr[\Gamma(\rho_i)W_j]$ is bounded by
\begin{equation}
    \label{eq:bound_meas}
    0 \leq \Delta^2_{(i,j)} \leq \frac{1}{d} \ .
\end{equation}
In order to estimate the $0-$fidelity, it is necessary to sample $l$ settings according to the probability distribution Eq.~\eqref{eq:prob_dist} and for each to estimate the corresponding $X(i,j)$ (Eq.~\eqref{eq:estimator}). This estimate $\hat{X}(i,j)$ is obtained using $m$ measurements with expected value $\mathds{E}[\hat{X}(i,j)]=X(i,j)$ and variance
\begin{equation}
\Delta^2 \hat{X}(i,j) = \frac{\Delta^2_{(i,j)}}{m \tr[\Lambda(\rho_i)W_j]^2} \ .
\end{equation}
We denote by $Y$ the random variable associated with such a sampling protocol. It follows a \textit{mixture distribution} with expected value $\mathds{E}[Y]=F_0$ and variance
\begin{align}
\label{eq:var_one}
\Delta^2Y &= \sum_{i,j} \operatorname{Pr}(i,j) \Big[\mathds{E}^2[\hat{X}(i,j)] + \Delta^2 \hat{X}(i,j)\Big] - \mathds{E}^2[Y] \\
&=\sum_{i,j} \left[ \frac{\tr[\Gamma(\rho_i)W_j]^2}{d^2} +  \frac{\Delta^2_{(i,j)}}{m d^2} \right] - F_0^2 \\
&= 1 + \frac{d^2 \Delta^2_{(i,j)}}{m} - F_0^2 \ ,
\end{align}
which can be bounded using Eq.~\eqref{eq:bound_meas} as
\begin{equation}
    1 - F_0^2 \leq \Delta^2Y \leq1 + d/m - F_0^2 \ .
\end{equation}
Estimates of the $0-$fidelity $\hat{F_0}$ may then be obtained by taking the mean of $l$ estimates $Y$, yielding a final variance of
\begin{equation}
    \label{eq:var_zerofid_est}
    0 \leq \frac{1 - F_0^2}{l} \leq \Delta^2\hat{F_0} \leq \frac{1 + d/m - F_0^2}{l} \leq \frac{1 + d/m}{l} \ ,
\end{equation}
where the outer bounds are obtained using $0\leq F_0 \leq 1$. From these bounds one can see that for a maximum number of settings $L$ (taken to be 900 in the main text) the variance is always greater than $(1-F_0^2)/L$ and saturates to this value in the limit $m\to \infty$. In the case where no restriction is placed on the number of unique settings, $m$ may be set to 1 and $l$ becomes the total number of measurements taken $N_e$ (that is, settings are sampled on a shot-by-shot basis). In the limit $N_e\to \infty$ both the upper and lower limits converge to 0 and thus no saturation to a non-zero value will be observed.

\subsection{Process Fidelity Estimates}
In order to estimate the process fidelity, the strategy presented in this work follows Ref.~\cite{flammia2011direct,da2011practical}: (i) Select a setting $(i,j)$, corresponding to a (non-directly) observable $\tr[\Lambda(\sigma_i)W_j]$, according to the probability distribution
\begin{equation}
\operatorname{Pr}'(i,j) = \frac{\tr[\Lambda(\sigma_i)W_j]^2}{d^2} \ ,
\end{equation}
(where the $\{\sigma_i\}$ and $\{W_j\}$ are both taken to be normalised Pauli operators) and (ii) for this setting, estimate the term
\begin{equation}
X'(i,j) = \frac{\tr[\Gamma(
\sigma_i)W_j]}{\tr[\Lambda(\sigma_i)W_j]}.
\end{equation}
In this case the estimate of the numerator ${\tr[\Gamma(
\sigma_i)W_j] = \sum_{k=1}^d \lambda^{\sigma_i}_k \tr[\Gamma(|\phi^{\sigma_i}_k\rangle\langle \phi^{\sigma_i}_k|) W_j]}$ is obtained based on $m'$ repeated measurements for each of the $d$ eigenvalues of $\sigma_i$. As with the $0-$fidelity, this gives rise to an estimate $\hat{X}'(i,j)$ of $X'(i,j)$ with expected value $\mathds{E}[\hat{X}'(i,j)]=X'(i,j)$ and variance
\begin{equation}
\Delta^2 \hat{X'}(i,j) = \frac{\Delta^2_{(i,j,k)}}{m' \tr[\Lambda(\sigma_i)W_j]^2} \ ,
\end{equation}
where $\Delta^2_{(i,j,k)}$ is the variance of a single measurement of a term ${\tr[\Gamma(|\phi^{\sigma_i}_k\rangle\langle \phi^{\sigma_i}_k|) W_j]}$, which is bounded (for the same reasoning as above) as
\begin{equation}
    0 \leq \Delta^2_{(i,j,k)} \leq \frac{1}{d} \ .
    \label{eq:bound_meas_2}
\end{equation}
We denote by $Y'$ the random variable associated with such a protocol, with expected value $\mathds{E}[Y']=F$ and variance
\begin{align}
\begin{split}
\label{eq:var_proc_2}
\Delta^2Y' &= \sum_{i,j} \operatorname{Pr}(i,j) \Big[\mathds{E}^2[\hat{X'}(i,j)] + \Delta^2 \hat{X'}(i,j)\Big] \\
& \quad \quad - \mathds{E}^2[Y'] \\
&=\sum_{i,j} \Big[ \frac{\tr^2[\Gamma(\sigma_i)W_j]}{d^2} +  \frac{\Delta^2_{(i,j,k)}}{m' d^2} \Big] - F^2 \\
&= 1 + \frac{d^2 \Delta^2_{(i,j,k)}}{m'} - F^2,
\end{split}
\end{align}
which can be bounded using Eq.~\ref{eq:bound_meas_2} as ${1 - F^2 \leq \Delta^2Y' \leq 1 + d/m' - F^2}$.
It follows that the estimate $\hat{F}$ of $F$ obtained over $l'$ sampled values of $Y'$ is unbiased with variance $\Delta^2\hat{F}$ bounded as
\begin{equation}
0 \leq \frac{1 - F^2}{l'} \leq \Delta^2\hat{F} \leq \frac{1 + d/m' - F^2}{l'} \leq \frac{1 + d/m'}{l'} \ .
\label{eq:bounds_process_2}
\end{equation}
where the outer bounds arise from the fact that $0\leq F\leq 1$. Although these bounds seem superficially similar to those obtained for the process fidelity, this neglects the fact that each of the $l'$ expectation values necessitates the measurement of $d$ unique circuits for a total number of experiments $dl'm'$.
In Figs.~\ref{fig:std_progression}~and~\ref{fig:toronto_statistics_evals} this is accounted for by setting $l'=l$ and $m'=m/d$ (where here $l$ and $m$ refer to the number of expectation values and measurement shots used in estimating the $0-$fidelity). This ensures that the total number of experiments $lm$ remains the same for process and $0-$fidelity estimations. This results in a modification of the bounds, yielding
\begin{equation}
0 \leq \frac{1 - F^2}{l} \leq \Delta^2\hat{F} \leq \frac{1 + d^2/m - F^2}{l} \leq \frac{1 + d^2/m}{l} \ .
\label{eq:bounds_process_manuscript}
\end{equation}

For a maximum of unique settings $L$ (corresponding to $L=ld$ to account for the need to input each of the $d$ eigenstates per setting $l$) the variance saturates to $d/(1-F^2)/L$ in the limit $m\to \infty$. In the case where there is no restriction on the number of unique settings, $m$ may be set to 1 and $l'$~(Eq.\eqref{eq:bounds_process_2}) effectively becomes the total number of measurements taken divided by the number of eigenstates, $l'=N_e/d$. In the limit $N_e\to \infty$ the variance goes to 0 and thus the saturation to a non-zero value would not be observed in this case.

\section{Optimality of the Input States}\label{ap:optimal_states}
The properties of the $0-$fidelity are dependent on the choice of input states $\{\rho_i\}$. As motivated in the main text, it is desirable to choose these states such that $\sum_{i\neq j}\tr[\rho_i \rho_j]$ is minimised whilst restricting to product states. In this section we show that the optimal choice of $n-$qubit product states is the tensor product of optimal single qubit states used in the main text.

An $n-$qubit product state may be denoted as ${\rho_{\vec{i}}=\rho^{(1)}_{i_1} \otimes \hdots \otimes \rho^{(n)}_{i_n}}$, where ${\vec{i}=[i_1,\hdots, i_n]}$ is the vector of index ${i_l \in \{1,2,3,4\}}$ labelling each of the single-qubit states $\rho^{(l)}_{i_l}$. Recalling that $\tr[A\otimes B]=\tr[A]\tr[B]$, the sum of overlaps between product states may be cast as
\begin{equation}
    \sum_{\vec{i}\neq \vec{j}} \tr[\rho_{\vec{i}}\rho_{\vec{j}}] = \sum_{\vec{i}\neq \vec{j}} \tr[\rho^{(1)}_{i_1}\rho^{(1)}_{j_1}] \tr[\rho^{(2)}_{i_2}\rho^{(2)}_{j_2}] \cdots \tr[\rho^{(n)}_{i_n}\rho^{(n)}_{j_n}] \ .
\end{equation}
For each pair of single qubit states $\rho^{(l)}_{i_l}, \rho^{(l)}_{j_l}$ this sum may be split into two separate sums, one where $\rho^{(l)}_{i_l} \neq \rho^{(l)}_{j_l}$ and one with $\rho^{(l)}_{i_l} = \rho^{(l)}_{j_l}$:
\begin{align}
    \sum_{\vec{i}\neq \vec{j}} \tr[\rho_{\vec{i}}\rho_{\vec{j}}] = &\sum_{i_l\neq j_l} \tr[\rho^{(l)}_{i_l} \rho^{(l)}_{j_l}]  \sum_{\vec{i^*},\vec{j^*}} \tr[\rho_{\vec{i^*}}\rho_{\vec{j^*}}] \nonumber \\
    &+ 4 \sum_{\vec{i^*} \neq\vec{j^*}} \tr[\rho_{\vec{i^*}}\rho_{\vec{j^*}}] \ ,
    \label{eq:trace_overlap}
\end{align}
where $\vec{i^*}$ represents the remaining vectors in $\vec{i}$ with the $l$th term removed. Each overlap between vector states $\tr[\rho_{\vec{i}^*}\rho_{\vec{j}^*}] \geq 0$; since it cannot be the case that all such overlaps are $0$ (since this would imply the states are orthogonal) the sum over all such states must be strictly positive, ${\sum_{\vec{i}^*\neq\vec{j}^*}\tr[\rho_{\vec{i}^*}\rho_{\vec{j}^*}]>0}$. It therefore follows that Eq.~\eqref{eq:trace_overlap} is minimised if and only if the sum of overlaps ${\sum_{i_l\neq j_l}\tr[\rho^{(l)}_{i_l}\rho^{(l)}_{j_l}]}$ is minimised \textit{i.e.} if the states $\{\rho^{(l)}_{i_l}\}$ are the single qubit SIC states given in the main text. Since this is true for all single qubit terms $l$ it necessarily follows that the tensor product of these states is the minimum over all possible tensor product states.

\section{Generation of Random Channels}\label{ap:rand_channels}
The numerical analyses of the process and $0-$fidelities in Secs.~\ref{sec:0_fidelity_properties} and~\ref{sec:estimation_comparisons} require the generation of random unitary channels. In this appendix the procedure for obtaining such channels is outlined.

\subsection{Random Unitary Channels}
In the numerical evaluations of the process and $0-$fidelities in Figs.~\ref{fig:B_expansion} and~\ref{fig:FOM_comparison} and in the full trace estimations presented in Fig.~\ref{fig:f_vs_FOM_dists}, the channels $\Lambda$ and $\Gamma$ were implemented by target ($U_t$) and comparison ($U_c$) unitary matrices as $\Lambda(\rho) = U_t \rho U^\dagger_t$ and $\Gamma(\rho) = U_c \rho U_c^\dagger$. Random target unitaries may be obtained by the exponentiation of random Hermitian matrices, which may themselves be generated as
\be
    H_t = \sum_{i_1,i_2,\cdots,i_n=0}^3 \alpha_{i_1i_2\cdots i_n} \bigotimes_{k=1}^n \sigma^{(i_k)}_k \ ,
    \label{eq:rand_herm}
\ee
where $\sigma_n^{(i_k)}$ are Pauli matrices acting on the $k-$th qubit (with $\sigma^{(0)}=\mathds{1}$) and where the coefficients $\alpha_{i_1i_2\cdots i_n}$ are sampled uniformly at random from the interval $[-1,1]$. With a random Hermitian matrix $H_t$ defined, the target unitary is then given by $U_t = e^{-i H_t}$.

The coefficients for the target unitary used in the numerical simulations of the $k-$fidelities presented in Fig.~\ref{fig:B_expansion} are given in Table~\ref{table:k_fidelity_target}. For the simulations of the $0-$fidelities shown in Fig.~\ref{fig:FOM_comparison}, the coefficients for the 2 and 3 qubit target unitaries are given in Tables~\ref{table:0_fidelity_target_2q} and~\ref{table:0_fidelity_target_3q} whilst those for the 5 qubit target unitary can be found in the GitHub repository~\cite{codeRepo}. The coefficients for the target unitary used in the full trace estimations presented in Fig.~\ref{fig:f_vs_FOM_dists} are given in Table~\ref{table:full_trace_ideal}.

For the comparison channels $U_c$, one could generate random unitary matrices in the same way, however for benchmarking purposes it is convenient to be able to control the fidelities of the evaluated pairs of channels. For this reason, the comparison unitaries $U_c$ were obtained as unitary rotations of the target unitary $U_t$ generated by a random Hermitian matrix $H_r$ as $U_c = e^{-i \epsilon H_r} U_t e^{i \epsilon H_r}$, where $\epsilon$ gives some control over the realised fidelities. For Figs.~\ref{fig:B_expansion} and~\ref{fig:FOM_comparison}, $\epsilon$ was varied from $0$ to $1$ to obtain evaluations over a full range of fidelities, whilst for Fig.~\ref{fig:f_vs_FOM_dists} a single value of $\epsilon=0.1$ was used. The coefficients for the random Hermitian matrix $H_r$ used in the full trace estimations presented in Fig.~\ref{fig:FOM_comparison} are given in Table~\ref{table:full_trace_rot}.

\subsection{Random Quantum Circuits}
\begin{figure}
  \includegraphics[width=\linewidth]{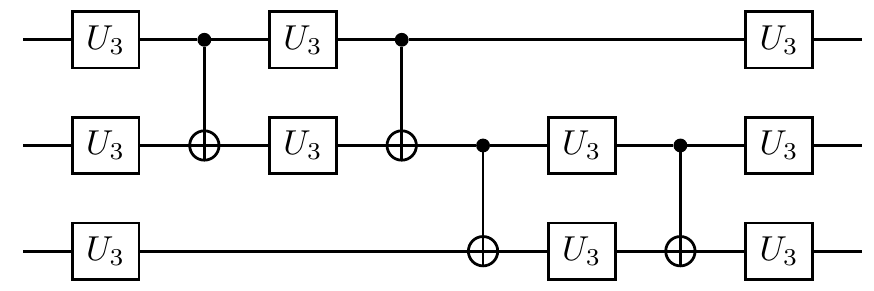}
  \caption{Circuit for the generation of random channels in Figs.~\ref{fig:std_progression} and~\ref{fig:toronto_statistics_evals}. Each of the $U_3$ gates has three parameters $\theta, \phi, \lambda$ which are randomly sampled from the interval $[0,2\pi]$.
  }
  \label{fig:random_circ}
\end{figure}

The projective estimations of the process and $0-$fidelities in Figs.~\ref{fig:std_progression} and~\ref{fig:toronto_statistics_evals} were performed using the circuit in Fig.~\ref{fig:random_circ} with all $30$ parameters sampled uniformly at random from the interval $[0,2\pi]$. The parameters correspond to 10 $U_3$ gates, each of which has three parameters $\theta, \phi, \lambda$ which define the gate as
\be
U_3(\theta,\phi,\lambda) =
\begin{pmatrix}
\cos{\theta/2} & -e^{i \lambda}\sin{\theta/2} \\
e^{i \phi}\sin{\theta/2} & e^{i(\lambda + \phi)}\cos{\theta/2}
\end{pmatrix} \ .
\ee
The specific parameters used are shown in Table~\ref{table:real_u3_params}. For the experimental estimations presented in Fig.~\ref{fig:toronto_statistics_evals} the target channel was obtained as the unitary representation of the circuit, and so any departure from the ideal fidelity arises from noise in the device.

For the numerical simulations presented in Fig.~\ref{fig:std_progression} the target channel was obtained as the unitary representation of the circuit and the comparison channel was generated by adding random coefficients sampled from the interval $[-0.4,0.4]$ to all of the $U_3$ parameters, with the same pair of channels being used for all numbers of total experiments. The specific parameters used in this work are given in Tables~\ref{table:num_ideal_u3_params} and~\ref{table:num_ideal_u3_params}. In these simulations various numbers of expectation values $l$ and shots $m$ were used; these experimental setups are given in Table~\ref{table:projective_exps_shots}.

\begin{table*}
\begin{center}
\begin{tabular}{|c|c|c|c|c|c|c|c|}
\hline
 $H_r$ term & $\alpha_{i_1i_2i_3}$ & $H_t$ term & $\alpha_{i_1i_2i_3}$ & $H_t$ term & $\alpha_{i_1i_2i_3}$ & $H_t$ term & $\alpha_{i_1i_2i_3}$ \\
 \hline
%  \hline
 $\sigma^{(0)}\sigma^{(0)}\sigma^{(0)}$ & 0.45631 & $\sigma^{(1)}\sigma^{(0)}\sigma^{(0)}$ & -0.70712 & $\sigma^{(2)}\sigma^{(0)}\sigma^{(0)}$ & -0.24903 & $\sigma^{(3)}\sigma^{(0)}\sigma^{(0)}$ & -0.55919 \\

 $\sigma^{(0)}\sigma^{(0)}\sigma^{(1)}$ & 0.52325 & $\sigma^{(1)}\sigma^{(0)}\sigma^{(1)}$ & 0.90678 & $\sigma^{(2)}\sigma^{(0)}\sigma^{(1)}$ & 0.41831 & $\sigma^{(3)}\sigma^{(0)}\sigma^{(1)}$ & 0.48434 \\

 $\sigma^{(0)}\sigma^{(0)}\sigma^{(2)}$ & 0.45163 & $\sigma^{(1)}\sigma^{(0)}\sigma^{(2)}$ & 0.33876 & $\sigma^{(2)}\sigma^{(0)}\sigma^{(2)}$ & 0.33918 & $\sigma^{(3)}\sigma^{(0)}\sigma^{(2)}$ & -0.10824 \\

 $\sigma^{(0)}\sigma^{(0)}\sigma^{(3)}$ & 0.07383 & $\sigma^{(1)}\sigma^{(0)}\sigma^{(3)}$ & 0.80992 & $\sigma^{(2)}\sigma^{(0)}\sigma^{(3)}$ & -0.96490 & $\sigma^{(3)}\sigma^{(0)}\sigma^{(3)}$ & 0.27820 \\

 $\sigma^{(0)}\sigma^{(1)}\sigma^{(0)}$ & 0.97577 & $\sigma^{(1)}\sigma^{(1)}\sigma^{(0)}$ & -0.89936 & $\sigma^{(2)}\sigma^{(1)}\sigma^{(0)}$ & -0.51887 & $\sigma^{(3)}\sigma^{(1)}\sigma^{(0)}$ & 0.855704 \\

 $\sigma^{(0)}\sigma^{(1)}\sigma^{(1)}$ & 0.11418 & $\sigma^{(1)}\sigma^{(1)}\sigma^{(1)}$ & -0.97843 & $\sigma^{(2)}\sigma^{(1)}\sigma^{(1)}$ & 0.81950 & $\sigma^{(3)}\sigma^{(1)}\sigma^{(1)}$ & 0.37004 \\

 $\sigma^{(0)}\sigma^{(1)}\sigma^{(2)}$ & -0.20607 & $\sigma^{(1)}\sigma^{(1)}\sigma^{(2)}$ & 0.23762 & $\sigma^{(2)}\sigma^{(1)}\sigma^{(2)}$ & -0.16243 & $\sigma^{(3)}\sigma^{(1)}\sigma^{(2)}$ & -0.66006 \\

 $\sigma^{(0)}\sigma^{(1)}\sigma^{(3)}$ & 0.45888 & $\sigma^{(1)}\sigma^{(1)}\sigma^{(3)}$ & 0.36063 & $\sigma^{(2)}\sigma^{(1)}\sigma^{(3)}$ & 0.56431 & $\sigma^{(3)}\sigma^{(1)}\sigma^{(3)}$ & 0.50428 \\

 $\sigma^{(0)}\sigma^{(2)}\sigma^{(0)}$ & -0.24559 & $\sigma^{(1)}\sigma^{(2)}\sigma^{(0)}$ & 0.31255 & $\sigma^{(2)}\sigma^{(2)}\sigma^{(0)}$ & -0.13876 & $\sigma^{(3)}\sigma^{(2)}\sigma^{(0)}$ & -0.90715 \\

 $\sigma^{(0)}\sigma^{(2)}\sigma^{(1)}$ & -0.00605 & $\sigma^{(1)}\sigma^{(2)}\sigma^{(1)}$ & 0.42754 & $\sigma^{(2)}\sigma^{(2)}\sigma^{(1)}$ & 0.32513 & $\sigma^{(3)}\sigma^{(2)}\sigma^{(1)}$ & -0.66542 \\

 $\sigma^{(0)}\sigma^{(2)}\sigma^{(2)}$ & 0.05597 & $\sigma^{(1)}\sigma^{(2)}\sigma^{(2)}$ & -0.29469 & $\sigma^{(2)}\sigma^{(2)}\sigma^{(2)}$ & 0.70243 & $\sigma^{(3)}\sigma^{(2)}\sigma^{(2)}$ & 0.05031 \\

 $\sigma^{(0)}\sigma^{(2)}\sigma^{(3)}$ & 0.37379 & $\sigma^{(1)}\sigma^{(2)}\sigma^{(3)}$ & -0.32656 & $\sigma^{(2)}\sigma^{(2)}\sigma^{(3)}$ & 0.14077 & $\sigma^{(3)}\sigma^{(2)}\sigma^{(3)}$ & 0.01150 \\

 $\sigma^{(0)}\sigma^{(3)}\sigma^{(0)}$ & 0.74952 & $\sigma^{(1)}\sigma^{(3)}\sigma^{(0)}$ & -0.49094 & $\sigma^{(2)}\sigma^{(3)}\sigma^{(0)}$ & -0.11147 & $\sigma^{(3)}\sigma^{(3)}\sigma^{(0)}$ & 0.47216 \\

 $\sigma^{(0)}\sigma^{(3)}\sigma^{(1)}$ & -0.98569 & $\sigma^{(1)}\sigma^{(3)}\sigma^{(1)}$ & -0.01268 & $\sigma^{(2)}\sigma^{(3)}\sigma^{(1)}$ & -0.13383 & $\sigma^{(3)}\sigma^{(3)}\sigma^{(1)}$ & 0.43830 \\

 $\sigma^{(0)}\sigma^{(3)}\sigma^{(2)}$ & 0.46017 & $\sigma^{(1)}\sigma^{(3)}\sigma^{(2)}$ & -0.70425 & $\sigma^{(2)}\sigma^{(3)}\sigma^{(2)}$ & 0.99265 & $\sigma^{(3)}\sigma^{(3)}\sigma^{(2)}$ & 0.57625 \\

 $\sigma^{(0)}\sigma^{(3)}\sigma^{(3)}$ & -0.58409 & $\sigma^{(1)}\sigma^{(3)}\sigma^{(3)}$ & -0.86759 & $\sigma^{(2)}\sigma^{(3)}\sigma^{(3)}$ & 0.07557 & $\sigma^{(3)}\sigma^{(3)}\sigma^{(3)}$ & 0.51732 \\
 \hline
\end{tabular}
\caption{Table of coefficients in the random Hermitian matrix (Eq.~\eqref{eq:rand_herm}) used to generate the random target unitary $U_t$ for the numerical evaluation of the $k-$fidelities presented in Fig.~\ref{fig:B_expansion}. \label{table:k_fidelity_target}}
\end{center}
\end{table*}

\begin{table*}
\begin{center}
\begin{tabular}{|c|c|c|c|}
\hline
 $H_t$ term & $\alpha_{i_1i_2}$ & $H_t$ term & $\alpha_{i_1i_2}$  \\
 \hline
 $\sigma^{(0)}\sigma^{(0)}$ & 0.62192 & $\sigma^{(2)}\sigma^{(0)}$ & 0.98298  \\
 $\sigma^{(0)}\sigma^{(1)}$ & -0.28442 & $\sigma^{(2)}\sigma^{(1)}$ & 0.15037  \\
 $\sigma^{(0)}\sigma^{(2)}$ & -0.36456 & $\sigma^{(2)}\sigma^{(2)}$ & -0.12910  \\
 $\sigma^{(0)}\sigma^{(3)}$ & -0.11006 & $\sigma^{(2)}\sigma^{(3)}$ & 0.78695  \\
 $\sigma^{(1)}\sigma^{(0)}$ & 0.13214 & $\sigma^{(3)}\sigma^{(0)}$ & -0.47983  \\
 $\sigma^{(1)}\sigma^{(1)}$ & -0.70606 & $\sigma^{(3)}\sigma^{(1)}$ & 0.30265  \\
 $\sigma^{(1)}\sigma^{(2)}$ & -0.66813 & $\sigma^{(3)}\sigma^{(2)}$ & -0.59174  \\
 $\sigma^{(1)}\sigma^{(3)}$ & -0.60901 & $\sigma^{(3)}\sigma^{(3)}$ & 0.33788  \\
 \hline
\end{tabular}
\caption{Table of coefficients in the random Hermitian matrix (Eq.~\eqref{eq:rand_herm}) used to generate the random target unitary $U_t$ for the numerical evaluation of the two qubit $0-$fidelities presented in Fig.~\ref{fig:FOM_comparison}. \label{table:0_fidelity_target_2q}}
\end{center}
\end{table*}

\begin{table*}
\begin{center}
\begin{tabular}{|c|c|c|c|c|c|c|c|}
\hline
 $H_r$ term & $\alpha_{i_1i_2i_3}$ & $H_t$ term & $\alpha_{i_1i_2i_3}$ & $H_t$ term & $\alpha_{i_1i_2i_3}$ & $H_t$ term & $\alpha_{i_1i_2i_3}$ \\
 \hline
%  \hline
 $\sigma^{(0)}\sigma^{(0)}\sigma^{(0)}$ & 0.17006 & $\sigma^{(1)}\sigma^{(0)}\sigma^{(0)}$ & -0.38200 & $\sigma^{(2)}\sigma^{(0)}\sigma^{(0)}$ & -0.84396 & $\sigma^{(3)}\sigma^{(0)}\sigma^{(0)}$ & 0.96076 \\

 $\sigma^{(0)}\sigma^{(0)}\sigma^{(1)}$ & 0.84389 & $\sigma^{(1)}\sigma^{(0)}\sigma^{(1)}$ & -0.87782 & $\sigma^{(2)}\sigma^{(0)}\sigma^{(1)}$ & 0.32516 & $\sigma^{(3)}\sigma^{(0)}\sigma^{(1)}$ & 0.69225 \\

 $\sigma^{(0)}\sigma^{(0)}\sigma^{(2)}$ & 0.77592 & $\sigma^{(1)}\sigma^{(0)}\sigma^{(2)}$ & -0.91405 & $\sigma^{(2)}\sigma^{(0)}\sigma^{(2)}$ & 0.28972 & $\sigma^{(3)}\sigma^{(0)}\sigma^{(2)}$ & 0.78857 \\

 $\sigma^{(0)}\sigma^{(0)}\sigma^{(3)}$ & -0.12891 & $\sigma^{(1)}\sigma^{(0)}\sigma^{(3)}$ & 0.85939 & $\sigma^{(2)}\sigma^{(0)}\sigma^{(3)}$ & 0.20802 & $\sigma^{(3)}\sigma^{(0)}\sigma^{(3)}$ & -0.97166 \\

 $\sigma^{(0)}\sigma^{(1)}\sigma^{(0)}$ & -0.66331 & $\sigma^{(1)}\sigma^{(1)}\sigma^{(0)}$ & 0.98379 & $\sigma^{(2)}\sigma^{(1)}\sigma^{(0)}$ & -0.78235 & $\sigma^{(3)}\sigma^{(1)}\sigma^{(0)}$ & 0.58005 \\

 $\sigma^{(0)}\sigma^{(1)}\sigma^{(1)}$ & 0.19222 & $\sigma^{(1)}\sigma^{(1)}\sigma^{(1)}$ & -0.70895 & $\sigma^{(2)}\sigma^{(1)}\sigma^{(1)}$ & -0.63625 & $\sigma^{(3)}\sigma^{(1)}\sigma^{(1)}$ & -0.93680 \\

 $\sigma^{(0)}\sigma^{(1)}\sigma^{(2)}$ & -0.00363 & $\sigma^{(1)}\sigma^{(1)}\sigma^{(2)}$ & -0.73675 & $\sigma^{(2)}\sigma^{(1)}\sigma^{(2)}$ & -0.86178 & $\sigma^{(3)}\sigma^{(1)}\sigma^{(2)}$ & -0.97715 \\

 $\sigma^{(0)}\sigma^{(1)}\sigma^{(3)}$ & -0.68100 & $\sigma^{(1)}\sigma^{(1)}\sigma^{(3)}$ & -0.96935 & $\sigma^{(2)}\sigma^{(1)}\sigma^{(3)}$ & 0.98716 & $\sigma^{(3)}\sigma^{(1)}\sigma^{(3)}$ & 0.34951 \\

 $\sigma^{(0)}\sigma^{(2)}\sigma^{(0)}$ & -0.32367 & $\sigma^{(1)}\sigma^{(2)}\sigma^{(0)}$ & 0.28944 & $\sigma^{(2)}\sigma^{(2)}\sigma^{(0)}$ & -0.78411 & $\sigma^{(3)}\sigma^{(2)}\sigma^{(0)}$ & 0.53184 \\

 $\sigma^{(0)}\sigma^{(2)}\sigma^{(1)}$ & -0.25096 & $\sigma^{(1)}\sigma^{(2)}\sigma^{(1)}$ & -0.98511 & $\sigma^{(2)}\sigma^{(2)}\sigma^{(1)}$ & 0.03250 & $\sigma^{(3)}\sigma^{(2)}\sigma^{(1)}$ & -0.02550 \\

 $\sigma^{(0)}\sigma^{(2)}\sigma^{(2)}$ & 0.00806 & $\sigma^{(1)}\sigma^{(2)}\sigma^{(2)}$ & -0.11461 & $\sigma^{(2)}\sigma^{(2)}\sigma^{(2)}$ & 0.07083 & $\sigma^{(3)}\sigma^{(2)}\sigma^{(2)}$ & 0.94426 \\

 $\sigma^{(0)}\sigma^{(2)}\sigma^{(3)}$ & 0.76286 & $\sigma^{(1)}\sigma^{(2)}\sigma^{(3)}$ & -0.22837 & $\sigma^{(2)}\sigma^{(2)}\sigma^{(3)}$ & 0.54588 & $\sigma^{(3)}\sigma^{(2)}\sigma^{(3)}$ & -0.02733 \\

 $\sigma^{(0)}\sigma^{(3)}\sigma^{(0)}$ & 0.71980 & $\sigma^{(1)}\sigma^{(3)}\sigma^{(0)}$ & 0.44338 & $\sigma^{(2)}\sigma^{(3)}\sigma^{(0)}$ & 0.16987 & $\sigma^{(3)}\sigma^{(3)}\sigma^{(0)}$ & 0.32453 \\

 $\sigma^{(0)}\sigma^{(3)}\sigma^{(1)}$ & 0.93134 & $\sigma^{(1)}\sigma^{(3)}\sigma^{(1)}$ & 0.18465 & $\sigma^{(2)}\sigma^{(3)}\sigma^{(1)}$ & -0.09857 & $\sigma^{(3)}\sigma^{(3)}\sigma^{(1)}$ & 0.92737 \\

 $\sigma^{(0)}\sigma^{(3)}\sigma^{(2)}$ & 0.33931 & $\sigma^{(1)}\sigma^{(3)}\sigma^{(2)}$ & 0.56110 & $\sigma^{(2)}\sigma^{(3)}\sigma^{(2)}$ & 0.09768 & $\sigma^{(3)}\sigma^{(3)}\sigma^{(2)}$ & 0.99687 \\

 $\sigma^{(0)}\sigma^{(3)}\sigma^{(3)}$ & 0.44647 & $\sigma^{(1)}\sigma^{(3)}\sigma^{(3)}$ & 0.35569 & $\sigma^{(2)}\sigma^{(3)}\sigma^{(3)}$ & -0.14916 & $\sigma^{(3)}\sigma^{(3)}\sigma^{(3)}$ & 0.74270 \\
 \hline
\end{tabular}
\caption{Table of coefficients in the random Hermitian matrix (Eq.~\eqref{eq:rand_herm}) used to generate the random target unitary $U_t$ for the numerical evaluation of the three qubit $0-$fidelities presented in Fig.~\ref{fig:FOM_comparison}. \label{table:0_fidelity_target_3q}}
\end{center}
\end{table*}

\begin{table*}
\begin{center}
\begin{tabular}{|c|c|c|c|c|c|c|c|}
\hline
 $H_t$ term & $\alpha_{i_1i_2i_3}$ & $H_t$ term & $\alpha_{i_1i_2i_3}$ & $H_t$ term & $\alpha_{i_1i_2i_3}$ & $H_t$ term & $\alpha_{i_1i_2i_3}$ \\
 \hline
 $\sigma^{(0)}\sigma^{(0)}\sigma^{(0)}$ & -0.12226 & $\sigma^{(1)}\sigma^{(0)}\sigma^{(0)}$ & 0.82557 & $\sigma^{(2)}\sigma^{(0)}\sigma^{(0)}$ & 0.07241 & $\sigma^{(3)}\sigma^{(0)}\sigma^{(0)}$ & -0.34315  \\

 $\sigma^{(0)}\sigma^{(0)}\sigma^{(1)}$ & -0.54535 & $\sigma^{(1)}\sigma^{(0)}\sigma^{(1)}$ & -0.24266 & $\sigma^{(2)}\sigma^{(0)}\sigma^{(1)}$ & 0.71791 & $\sigma^{(3)}\sigma^{(0)}\sigma^{(1)}$ & -0.60957 \\

 $\sigma^{(0)}\sigma^{(0)}\sigma^{(2)}$ & 0.07770 & $\sigma^{(1)}\sigma^{(0)}\sigma^{(2)}$ & 0.83829 & $\sigma^{(2)}\sigma^{(0)}\sigma^{(2)}$ & 0.12453 & $\sigma^{(3)}\sigma^{(0)}\sigma^{(2)}$ & 0.07668 \\

 $\sigma^{(0)}\sigma^{(0)}\sigma^{(3)}$ & -0.83056 & $\sigma^{(1)}\sigma^{(0)}\sigma^{(3)}$ & -0.92854 & $\sigma^{(2)}\sigma^{(0)}\sigma^{(3)}$ & 0.35785 & $\sigma^{(3)}\sigma^{(0)}\sigma^{(3)}$ & -0.09843 \\

 $\sigma^{(0)}\sigma^{(1)}\sigma^{(0)}$ & -0.44200 & $\sigma^{(1)}\sigma^{(1)}\sigma^{(0)}$ & -0.22618 & $\sigma^{(2)}\sigma^{(1)}\sigma^{(0)}$ & 0.43885 & $\sigma^{(3)}\sigma^{(1)}\sigma^{(0)}$ & -0.99722 \\

 $\sigma^{(0)}\sigma^{(1)}\sigma^{(1)}$ & -0.65783 & $\sigma^{(1)}\sigma^{(1)}\sigma^{(1)}$ & 0.36146 & $\sigma^{(2)}\sigma^{(1)}\sigma^{(1)}$ & -0.93251 & $\sigma^{(3)}\sigma^{(1)}\sigma^{(1)}$ & -0.61818 \\

 $\sigma^{(0)}\sigma^{(1)}\sigma^{(2)}$ & -0.72367 & $\sigma^{(1)}\sigma^{(1)}\sigma^{(2)}$ & 0.17789 & $\sigma^{(2)}\sigma^{(1)}\sigma^{(2)}$ & 0.45259 & $\sigma^{(3)}\sigma^{(1)}\sigma^{(2)}$ & -0.91497 \\

 $\sigma^{(0)}\sigma^{(1)}\sigma^{(3)}$ & -0.06722 & $\sigma^{(1)}\sigma^{(1)}\sigma^{(3)}$ & 0.13096 & $\sigma^{(2)}\sigma^{(1)}\sigma^{(3)}$ & -0.26926 & $\sigma^{(3)}\sigma^{(1)}\sigma^{(3)}$ & 0.75595 \\

 $\sigma^{(0)}\sigma^{(2)}\sigma^{(0)}$ & -0.75042 & $\sigma^{(1)}\sigma^{(2)}\sigma^{(0)}$ & -0.78464 & $\sigma^{(2)}\sigma^{(2)}\sigma^{(0)}$ & -0.30934 & $\sigma^{(3)}\sigma^{(2)}\sigma^{(0)}$ & 0.71169  \\

 $\sigma^{(0)}\sigma^{(2)}\sigma^{(1)}$ & -0.75602 & $\sigma^{(1)}\sigma^{(2)}\sigma^{(1)}$ & 0.76222 & $\sigma^{(2)}\sigma^{(2)}\sigma^{(1)}$ & 0.79482 & $\sigma^{(3)}\sigma^{(2)}\sigma^{(1)}$ & 0.04450 \\

 $\sigma^{(0)}\sigma^{(2)}\sigma^{(2)}$ & 0.67848 & $\sigma^{(1)}\sigma^{(2)}\sigma^{(2)}$ & 0.50086 & $\sigma^{(2)}\sigma^{(2)}\sigma^{(2)}$ & 0.56477 & $\sigma^{(3)}\sigma^{(2)}\sigma^{(2)}$ & -0.22242 \\

 $\sigma^{(0)}\sigma^{(2)}\sigma^{(3)}$ & -0.99924 & $\sigma^{(1)}\sigma^{(2)}\sigma^{(3)}$ & -0.80395 & $\sigma^{(2)}\sigma^{(2)}\sigma^{(3)}$ & 0.20221 & $\sigma^{(3)}\sigma^{(2)}\sigma^{(3)}$ & 0.88407 \\

 $\sigma^{(0)}\sigma^{(3)}\sigma^{(0)}$ & 0.03075 & $\sigma^{(1)}\sigma^{(3)}\sigma^{(0)}$ & 0.30451 & $\sigma^{(2)}\sigma^{(3)}\sigma^{(0)}$ & 0.88503 & $\sigma^{(3)}\sigma^{(3)}\sigma^{(0)}$ & -0.23753 \\

 $\sigma^{(0)}\sigma^{(3)}\sigma^{(1)}$ & -0.95405 & $\sigma^{(1)}\sigma^{(3)}\sigma^{(1)}$ & -0.90343 & $\sigma^{(2)}\sigma^{(3)}\sigma^{(1)}$ & -0.93316 & $\sigma^{(3)}\sigma^{(3)}\sigma^{(1)}$ & 0.32873 \\

 $\sigma^{(0)}\sigma^{(3)}\sigma^{(2)}$ & -0.42508 & $\sigma^{(1)}\sigma^{(3)}\sigma^{(2)}$ & -0.66899 & $\sigma^{(2)}\sigma^{(3)}\sigma^{(2)}$ & -0.77155 & $\sigma^{(3)}\sigma^{(3)}\sigma^{(2)}$ & 0.77092 \\

 $\sigma^{(0)}\sigma^{(3)}\sigma^{(3)}$ & -0.31457 & $\sigma^{(1)}\sigma^{(3)}\sigma^{(3)}$ & -0.19482 & $\sigma^{(2)}\sigma^{(3)}\sigma^{(3)}$ & -0.92350 & $\sigma^{(3)}\sigma^{(3)}\sigma^{(3)}$ & -0.54485 \\
 \hline
\end{tabular}
\caption{Table of coefficients in the random Hermitian matrix (Eq.~\eqref{eq:rand_herm}) used to generate the random target unitary $U_t$ for the numerical evaluation of the full trace process and $0-$fidelity estimations presented in Fig.~\ref{fig:f_vs_FOM_dists}. \label{table:full_trace_ideal}}
\end{center}
\end{table*}

\begin{table*}
\begin{center}
\begin{tabular}{|c|c|c|c|c|c|c|c|}
\hline
 $H_r$ term & $\alpha_{i_1i_2i_3}$ & $H_r$ term & $\alpha_{i_1i_2i_3}$ & $H_r$ term & $\alpha_{i_1i_2i_3}$ & $H_r$ term & $\alpha_{i_1i_2i_3}$ \\
 \hline
%  \hline
 $\sigma^{(0)}\sigma^{(0)}\sigma^{(0)}$ & 0.50016 & $\sigma^{(1)}\sigma^{(0)}\sigma^{(0)}$ & 0.23656 & $\sigma^{(2)}\sigma^{(0)}\sigma^{(0)}$ & -0.16832 & $\sigma^{(3)}\sigma^{(0)}\sigma^{(0)}$ & -0.94659  \\

 $\sigma^{(0)}\sigma^{(0)}\sigma^{(1)}$ & 0.81491 & $\sigma^{(1)}\sigma^{(0)}\sigma^{(1)}$ & 0.02643 & $\sigma^{(2)}\sigma^{(0)}\sigma^{(1)}$ & 0.81125 & $\sigma^{(3)}\sigma^{(0)}\sigma^{(1)}$ & 0.59747 \\

 $\sigma^{(0)}\sigma^{(0)}\sigma^{(2)}$ & -0.17869 & $\sigma^{(1)}\sigma^{(0)}\sigma^{(2)}$ & 0.49818 & $\sigma^{(2)}\sigma^{(0)}\sigma^{(2)}$ & -0.01940 & $\sigma^{(3)}\sigma^{(0)}\sigma^{(2)}$ & 0.93269 \\

 $\sigma^{(0)}\sigma^{(0)}\sigma^{(3)}$ & -0.30403 & $\sigma^{(1)}\sigma^{(0)}\sigma^{(3)}$ & 0.74331 & $\sigma^{(2)}\sigma^{(0)}\sigma^{(3)}$ & -0.07383 & $\sigma^{(3)}\sigma^{(0)}\sigma^{(3)}$ & -0.55021 \\

 $\sigma^{(0)}\sigma^{(1)}\sigma^{(0)}$ & -0.79330 & $\sigma^{(1)}\sigma^{(1)}\sigma^{(0)}$ & -0.27526 & $\sigma^{(2)}\sigma^{(1)}\sigma^{(0)}$ & 0.69006 & $\sigma^{(3)}\sigma^{(1)}\sigma^{(0)}$ & 0.92612 \\

 $\sigma^{(0)}\sigma^{(1)}\sigma^{(1)}$ & -0.85750 & $\sigma^{(1)}\sigma^{(1)}\sigma^{(1)}$ & -0.30359 & $\sigma^{(2)}\sigma^{(1)}\sigma^{(1)}$ & -0.02089 & $\sigma^{(3)}\sigma^{(1)}\sigma^{(1)}$ & 0.76278 \\

 $\sigma^{(0)}\sigma^{(1)}\sigma^{(2)}$ & 0.32984 & $\sigma^{(1)}\sigma^{(1)}\sigma^{(2)}$ & -0.47171 & $\sigma^{(2)}\sigma^{(1)}\sigma^{(2)}$ & 0.95378 & $\sigma^{(3)}\sigma^{(1)}\sigma^{(2)}$ & -0.42033 \\

 $\sigma^{(0)}\sigma^{(1)}\sigma^{(3)}$ & 0.50314 & $\sigma^{(1)}\sigma^{(1)}\sigma^{(3)}$ & -0.99054 & $\sigma^{(2)}\sigma^{(1)}\sigma^{(3)}$ & 0.30379 & $\sigma^{(3)}\sigma^{(1)}\sigma^{(3)}$ & -0.00098 \\

 $\sigma^{(0)}\sigma^{(2)}\sigma^{(0)}$ & -0.85538 & $\sigma^{(1)}\sigma^{(2)}\sigma^{(0)}$ & 0.11111 & $\sigma^{(2)}\sigma^{(2)}\sigma^{(0)}$ & 0.24740 & $\sigma^{(3)}\sigma^{(2)}\sigma^{(0)}$ & 0.89360  \\

 $\sigma^{(0)}\sigma^{(2)}\sigma^{(1)}$ & -0.94635 & $\sigma^{(1)}\sigma^{(2)}\sigma^{(1)}$ & 0.78359 & $\sigma^{(2)}\sigma^{(2)}\sigma^{(1)}$ & 0.54317 & $\sigma^{(3)}\sigma^{(2)}\sigma^{(1)}$ & 0.32564 \\

 $\sigma^{(0)}\sigma^{(2)}\sigma^{(2)}$ & 0.37845 & $\sigma^{(1)}\sigma^{(2)}\sigma^{(2)}$ & -0.46577 & $\sigma^{(2)}\sigma^{(2)}\sigma^{(2)}$ & -0.44707 & $\sigma^{(3)}\sigma^{(2)}\sigma^{(2)}$ & -0.73448 \\

 $\sigma^{(0)}\sigma^{(2)}\sigma^{(3)}$ & -0.77559 & $\sigma^{(1)}\sigma^{(2)}\sigma^{(3)}$ & -0.36554 & $\sigma^{(2)}\sigma^{(2)}\sigma^{(3)}$ & -0.67578 & $\sigma^{(3)}\sigma^{(2)}\sigma^{(3)}$ & 0.28747 \\

 $\sigma^{(0)}\sigma^{(3)}\sigma^{(0)}$ & -0.59413 & $\sigma^{(1)}\sigma^{(3)}\sigma^{(0)}$ & -0.42969 & $\sigma^{(2)}\sigma^{(3)}\sigma^{(0)}$ & 0.44714 & $\sigma^{(3)}\sigma^{(3)}\sigma^{(0)}$ & -0.98854 \\

 $\sigma^{(0)}\sigma^{(3)}\sigma^{(1)}$ & 0.86622 & $\sigma^{(1)}\sigma^{(3)}\sigma^{(1)}$ & 0.27964 & $\sigma^{(2)}\sigma^{(3)}\sigma^{(1)}$ & -0.95976 & $\sigma^{(3)}\sigma^{(3)}\sigma^{(1)}$ & 0.08694 \\

 $\sigma^{(0)}\sigma^{(3)}\sigma^{(2)}$ & 0.11537 & $\sigma^{(1)}\sigma^{(3)}\sigma^{(2)}$ & 0.97530 & $\sigma^{(2)}\sigma^{(3)}\sigma^{(2)}$ & 0.69712 & $\sigma^{(3)}\sigma^{(3)}\sigma^{(2)}$ & -0.38816 \\

 $\sigma^{(0)}\sigma^{(3)}\sigma^{(3)}$ & 0.66967 & $\sigma^{(1)}\sigma^{(3)}\sigma^{(3)}$ & 0.34516 & $\sigma^{(2)}\sigma^{(3)}\sigma^{(3)}$ & -0.35423 & $\sigma^{(3)}\sigma^{(3)}\sigma^{(3)}$ & -0.18427 \\
 \hline
\end{tabular}
\caption{Table of coefficients in the random Hermitian matrix (Eq.~\eqref{eq:rand_herm}) used to generate the random comparison unitary $e^{-iH_r}$ for the numerical evaluation of the full trace process and $0-$fidelity estimations presented in Fig.~\ref{fig:f_vs_FOM_dists}. \label{table:full_trace_rot}}
\end{center}
\end{table*}

\begin{table*}
\begin{center}
\begin{tabular}{|c|c|c|c|}
\hline
 $U_3$ gate & $\theta$ & $\phi$ & $\lambda$  \\
 \hline
%  \h
 $U_3^{(1)}$ & 4.64699 & 5.16852 & 4.38670  \\
 $U_3^{(2)}$ & 5.04437 & 0.62442 & 4.59349  \\
 $U_3^{(3)}$ & 5.89901 & 3.90661 & 6.21039  \\
 $U_3^{(4)}$ & 1.52430 & 1.40088 & 6.07413  \\
 $U_3^{(5)}$ & 0.12721 & 0.36008 & 6.12182  \\
 $U_3^{(6)}$ & 1.37419 & 3.14458 & 5.41173  \\
 $U_3^{(7)}$ & 5.18368 & 0.12673 & 1.91885  \\
 $U_3^{(8)}$ & 4.74807 & 2.12120 & 5.92042  \\
 $U_3^{(9)}$ & 3.86544 & 5.88484 & 0.75772  \\
 $U_3^{(10)}$ & 5.64946 & 0.37099 & 2.58326  \\
 \hline
\end{tabular}
\caption{Table of $U_3$ parameters used in the circuit in Fig.~\ref{fig:random_circ} for the generation of a random benchmarking channel. The channel was implemented on the \texttt{ibmq\_toronto} quantum device, with all deviations from ideality arising from noise in the machine. \label{table:real_u3_params}}
\end{center}
\end{table*}

\begin{table*}
\begin{center}
\begin{tabular}{|c|c|c|c|}
\hline
 $U_3$ gate & $\theta$ & $\phi$ & $\lambda$  \\
 \hline
 $U_3^{(1)}$ & 4.84482 & 4.76108 & 5.94502  \\
 $U_3^{(2)}$ & 0.39148 & 3.84468 & 5.47668  \\
 $U_3^{(3)}$ & 5.75781 & 1.08062 & 5.21778  \\
 $U_3^{(4)}$ & 0.75037 & 4.38013 & 0.00859  \\
 $U_3^{(5)}$ & 1.23687 & 6.01724 & 1.93567  \\
 $U_3^{(6)}$ & 1.20877 & 0.87278 & 1.85682  \\
 $U_3^{(7)}$ & 2.64055 & 3.00598 & 1.23498  \\
 $U_3^{(8)}$ & 1.00054 & 5.32563 & 0.46313  \\
 $U_3^{(9)}$ & 3.63063 & 2.04242 & 5.88751 \\
 $U_3^{(10)}$ & 4.86354 & 2.56637 & 3.34799  \\
 \hline
\end{tabular}
\caption{Table of $U_3$ parameters used in the circuit in Fig.~\ref{fig:random_circ} for the generation of a target unitary $U_t$. \label{table:num_ideal_u3_params}}
\end{center}
\end{table*}

\begin{table*}
\begin{center}
\begin{tabular}{|c|c|c|c|}
\hline
 $U_3$ gate & $\theta$ & $\phi$ & $\lambda$  \\
 \hline
 $U_3^{(1)}$ & 4.81392 & 4.45876 & 5.75332  \\
 $U_3^{(2)}$ & 0.49220 & 3.48554 & 5.51691  \\
 $U_3^{(3)}$ & 5.79802 & 0.81921 & 5.60671  \\
 $U_3^{(4)}$ & 0.43546 & 4.76768 & -0.15634  \\
 $U_3^{(5)}$ & 0.98126 & 6.13769 & 1.65627  \\
 $U_3^{(6)}$ & 1.43111 & 0.82898 & 1.90723  \\
 $U_3^{(7)}$ & 2.39113 & 3.02910 & 1.32447  \\
 $U_3^{(8)}$ & 0.62534 & 5.26572 & 0.35299  \\
 $U_3^{(9)}$ & 3.61569 & 2.20167 & 5.89594 \\
 $U_3^{(10)}$ & 4.86430 & 2.60815 & 3.67214  \\
 \hline
\end{tabular}
\caption{Table of $U_3$ parameters used in the circuit in Fig.~\ref{fig:random_circ} for the generation of the comparison unitary. \label{table:num_compare_u3_params}}
\end{center}
\end{table*}

\begin{table*}
\begin{center}
\begin{tabular}{|c|c|c|c|c|c|}
 \cline{2-6}
 \multicolumn{1}{c|}{}& \multicolumn{2}{|c|}{$0-$fidelity} & \multicolumn{3}{|c|}{Process fidelity} \\
 \hline
 Total Experiments $lm$ & Expectation values $l$ & Shots $m$ & Expectation values $l$ & Unique experiments $dl$ & Shots $m$ \\
 \hline
 896 & 28 & 32 & 28 & 224 & 4 \\
 3584 & 56 & 64 & 56 & 448 & 8 \\
 14336 & 112 & 128 & 112 & 896 & 16 \\
 57344 & 224 & 256 & 112 & 896 & 64 \\
 129024 & 336 & 384 & 112 & 896 & 144 \\
 229376 & 448 & 512 & 112 & 896 & 256 \\
 358400 & 560 & 640 & 112 & 896 & 400 \\
 516096 & 672 & 768 & 112 & 896 & 576 \\
 702464 & 784 & 896 & 112 & 896 & 784 \\
 917504 & 896 & 1024 & 112 & 896 & 1024 \\
 \hline
\end{tabular}
\caption{Table showing the number of experiments $l$ and the number of shots $m$ for each set of total experiments $lm$ assessed in the numerical evaluation of the projective process and $0-$fidelity estimations presented in Fig.~\ref{fig:std_progression}. \label{table:projective_exps_shots}}
\end{center}
\end{table*}

\bibliography{bibliography.bib}

\end{document}